\documentclass[usenatbib]{mnras}

\usepackage[T1]{fontenc}
\usepackage{ae,aecompl}

\usepackage{graphicx}	
\usepackage{amsmath}	
\usepackage{amssymb}	
\usepackage{booktabs}
\usepackage{eso-pic}

\usepackage{ulem}       

\usepackage{times}

\newcommand{\Planck}{{\slshape Planck~}}

\def \cf2{{\it Cosmicflows-2\,}}

\def \der{{\rm d}}
\def \msun{{{\rm M}_{\odot}}}

\def \araa{Annual Review of Astron and Astrophys}
\def\mnras{Monthly Notices of the Royal Astronomical Society}

\def\aap{Astronomy and Astrophysics}

\def\jcap{Journal of Cosmology and Astroparticle Physics}

\makeatletter
\let\ftype@table\ftype@figure
\makeatother

\title[Gas physics with LRGs]{Probing galaxy cluster and intra-cluster gas with luminous red galaxies}
\author[Gong, Ma, \& Tanimura]{Yan Gong$^{1,3}$, Yin-Zhe Ma$^{2,3}$, and Hideki Tanimura$^4$
\\
$^{1}$ Key Laboratory for Computational Astrophysics, National Astronomical Observatories, Chinese Academy of Sciences, \\20A Datun Road, Beijing 100012, China\\
$^{2}$ School of Chemistry and Physics, University of KwaZulu-Natal, Westville Campus, Private Bag X54001, Durban, 4000, South Africa \\
$^{3}$ NAOC-UKZN Computational Astrophysics Centre (NUCAC), University of KwaZulu-Natal, Durban, 4000, South Africa \\
$^{4}$ Institut d'Astrophysique Spatiale, CNRS (UMR 8617), Université Paris-Sud, Bâtiment 121, Orsay, France
}

\pubyear{2018}

\begin{document}
\label{firstpage}
\pagerange{\pageref{firstpage}--\pageref{lastpage}}
\maketitle

\begin{abstract}
We use the cross-correlation between the thermal Sunyaev-Zeldovich  (tSZ) signal measured by the {\it Planck} satellite and the luminous red galaxy (LRG) samples provided by the SDSS DR7 to study the properties of galaxy cluster and intra-cluster gas. We separate the samples into three redshift bins $z_{1}=(0.16,\,0.26)$, $z_{2}=(0.26,\,0.36)$, $z_{3}=(0.36,\,0.47)$, and  stack the {\it Planck} $y$-map against LRGs to derive the averaged $y$ profile for each redshift bin. We then fit the stacked profile with the theoretical prediction from the  universal pressure profile (UPP) by using the Markov-Chain Monte-Carlo method. We find that the best-fit values of the UPP parameters for the three bins are generally consistent with the previous studies, except for the noticeable evolution of the parameters in the three redshift bins. We simultaneously fit the data in the three redshift bins together, and find that the original UPP model cannot fit the data at small angular scales very well in the first and third redshift bins. The joint fits can be improved by including an additional parameter $\eta$ to change the redshift-dependence of the model (i.e. $E(z)^{8/3}\, \rightarrow \,E(z)^{8/3+\eta}$) with best-fit value as $\eta=-3.11^{+1.09}_{-1.13}$. This suggests that the original UPP model with less redshift-dependence may provide a better fit to the stacked thermal Sunyaev-Zeldovich profile.
\end{abstract}
%

%



\section{Introduction}
\label{sec:intro}
The clusters of galaxies are the important objects in learning the structure formation and cosmology in general. Around hot clusters, the gas is generally ionized with temperature above $10^{6}\,$K~\citep{McCarthy07} and provides high-density of hot electrons. The intra-cluster medium (ICM) can be studied by using direct X-ray imaging or the Sunyaev-Zeldovich effect (SZ effect;~\citealt{Sunyaev72,Sunyaev80}). Intensity of the former is proportional to the square of the electron density $\sim n^{2}_{\rm e}$ which could be sensitive to the center baryons. In contrast, observation of the SZ effect depends on the integration of pressure profile, which could be used to trace down the gas distribution in the lower density region. These regions, such as filament, sheets and voids, are the locations believed to host most of the baryons~\citep{Haider16,Tanimura19a}. For this reason, there has been a growing interest in predicting and measuring the SZ effect by using various techniques in microwave and radio bands~\citep{Birkinshaw78,Birkinshaw99,Carlstrom02,Ma15}. 

The thermal SZ (tSZ) effect is a secondary anisotropy in the cosmic microwave background radiation (CMB), which is caused by the hot free electrons around ICM scattering off the CMB photons, boosting the low-frequency CMB into higher frequency, i.e.
\begin{eqnarray}
\frac{\Delta T}{T_{\rm CMB}}
= \left[\eta \frac{e^{\eta}+1}{e^{\eta}-1} - 4\right] y \equiv
g_{\nu}y, \label{deltaT1}
\end{eqnarray} 
where $g_{\nu} \equiv
[\eta(e^{\eta}+1)/(e^{\eta}-1)] - 4$
captures the frequency
dependence, and
\begin{eqnarray}
\eta = \frac{h \nu}{k_{\rm{B}}T_{\rm{CMB}}} =
1.76 \left(\frac{\nu}{100 \textrm{ GHz}}\right). \label{xdefine}
\end{eqnarray}
The dimensionless Comptonization parameter $y$ is
\begin{eqnarray}
y = \int n_{\rm{e}}({\bf
r})\sigma_{\rm{T}} \,\frac{k_{\rm{B}} T_{\rm{e}}({\bf
r})}{m_{\rm{e}} c^2} \,\der l, \label{comptony}
\end{eqnarray}   
where
$\sigma_{\rm{T}}$ is the Thomson cross section, $k_{\rm{B}}$ is
the Boltzmann constant, $m_{\rm{e}}$ is the electron rest mass and
the integral is taken along the line of sight. The sign of $g_{\nu}$ determines whether $\Delta T$ is positive or negative (an increment or decrement) for the CMB temperature.  With $T_{\rm CMB} = 2.725$ K, we have $g_{\nu} \geq 0$ for $\nu \geq 217$ GHz, and vice versa. We can define the electron pressure profile as $P_{\rm e}({\bf r})=n_{\rm e}({\bf r})\,k_{\rm B}T_{\rm e}({\bf r})$ in Eq.~(\ref{comptony}), and thus the pressure profile $P_{\rm e}$ is an essential factor in determining the Compton $y$-parameter. Therefore modelling and testing the pressure profile is very crucial to obtain an accurate profile of observed $y$-parameter. 

One the other hand, by accurately measuring the Compton-$y$ parameter, a more precise modelling of the pressure profile can be obtained. In 2010, \citet{Arnaud10} investigated the cluster pressure profile with $33$ local ($z<0.2$) samples of clusters drawn from {\tt REXCESS} catalogue observed by {\it XMM-Newton}. These samples span the mass range of $10^{14}\msun<M_{500}<10^{15}\msun$ where the $M_{500}$ is the mass of the center of cluster within the radius of density equal to $500\rho_{\rm crit}$. \citet{Arnaud10} proposed the {\it Universal Pressure Profile} (UPP) for the clusters and determined the parameter values by using the $33$ massive clusters. Following~\citet{Arnaud10}, \citet{Planck13} used the {\it Planck}'s 14-month nominal survey and $62$ massive nearby clusters, and re-investigated the cluster UPP, and refined the parameter values. Besides these observational probes, \citet{LeBrun15} used the cosmo-OWLS simulation and tested the UPP against different AGN models, and refined the UPP model parameters.

In these studies, the tSZ effect is a very useful tool to explore the high-redshift galaxy clusters. Because the signal $y$ is expected to be nearly independent of the redshift, the effect does not diminish with the increasing redshift. Therefore the effect is very suitable for finding high-redshift clusters~\citep{Planck11}, and cross-correlation with other large-scale structure tracers (e.g. cross-correlation with weak lensing measurement~\citep{Ma15,Hojjati17}). In addition, luminous red galaxies (LRGs) are early-type massive galaxies consisting mainly of old stars with little ongoing star formation. Thus LRGs are good tracers of galaxy clusters and underlying dark matter distribution, and should be correlated with thermal SZ signal~\citep{Hoshino2015}. 

In this work, we will study this cross-correlation and explore the UPP of clusters. Since the Sloan Digital Sky Survey (SDSS) released its data, there have been increasingly large LRG catalogues that span more orders of magnitude in mass and occupy different distances. The samples we use here is DR7 LRG samples, which has mass range $3\times10^{12}\msun$ to $3\times10^{14}\msun$ and redshift range $0.16<z<0.47$, and significantly overlap with the {\it Planck} all-sky Compton $y-$map. Therefore, providing the stacking results of the LRG samples at cosmological distances and fitting the UPP model is the main aim of this paper. We will provide robust tests of the UPP model parameters by this large LRG sample from SDSS DR7.
  
This paper is organised as follows. In Sec.~\ref{sec:data}, we present the {\it Planck} $y$-maps from the CMB and LRG catalogues from SDSS-DR7 used in this study, and the stacking results of LRG $y$-profile. In Sec.~\ref{sec:model}, we present the modelling of the stacked profiles by calculating the 1-halo and 2-halo terms of the stacked profile. In Sec.~\ref{sec:analysis}, we discuss the data analysis procedure and $\chi^{2}$ study in our work. In Sec.~\ref{sec:result}, we present the results of our numerical fitting and compare our results with other works. The summary and discussion are presented in the last section.

Throughout this paper, we adopt a spatially-flat fiducial $\Lambda$CDM cosmology with {\it Planck} cosmological parameter values~\citep{Planck13-parameters}, which are the fractional baryon density $\Omega_{\rm b}=0.0490$, fractional matter density $\Omega_{\rm m}=0.3175$, spectral index $n_{\rm s}=0.9624$, {\it rms} matter fluctuation amplitude $\sigma_{8}=0.834$, and $h=0.6711$ defined as $H_{0}=100\,h\,{\rm km}\,{\rm s}^{-1}\,{\rm Mpc}^{-1}$, where $H_0$ is the Hubble constant. We note that our constraint results are not very sensitive to the current determined values of cosmological parameters.

\begin{figure}
\includegraphics[scale = 0.41]{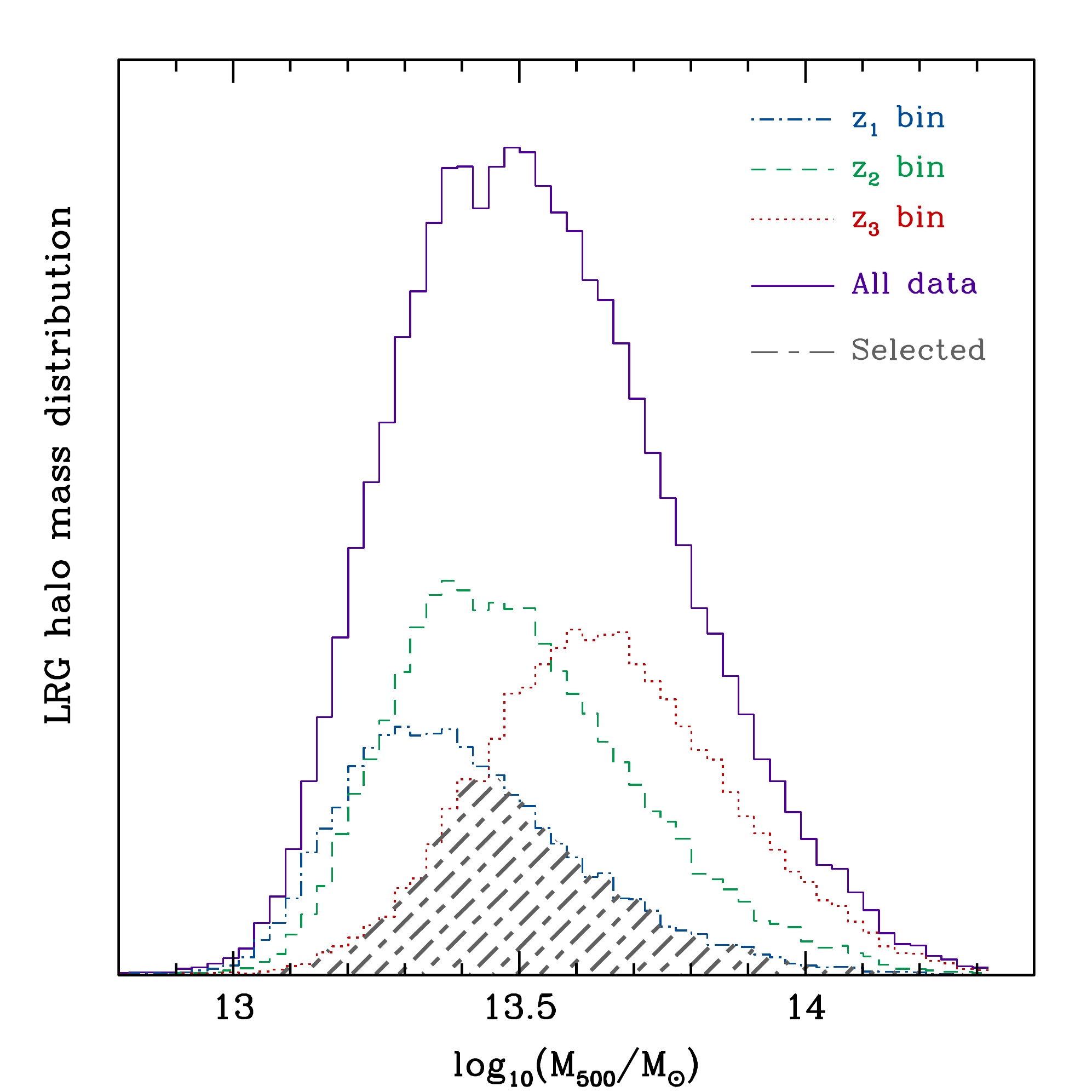}
\caption{\label{fig:M_dis} The halo mass distributions of the LRG samples in different redshift ranges. The blue, green, and red dashed curves are for $z_1$, $z_2$, and $z_3$ bins, respectively. The mass distribution of the whole redshift range is denoted in solid purple curve. The gray hatched region shows the selected sub-samples, which are used in our analysis, with the same mass distribution for the three $z$-bins.}
\end{figure}

\begin{figure*}
\centerline{
\resizebox{!}{!}{\includegraphics[scale=0.4]{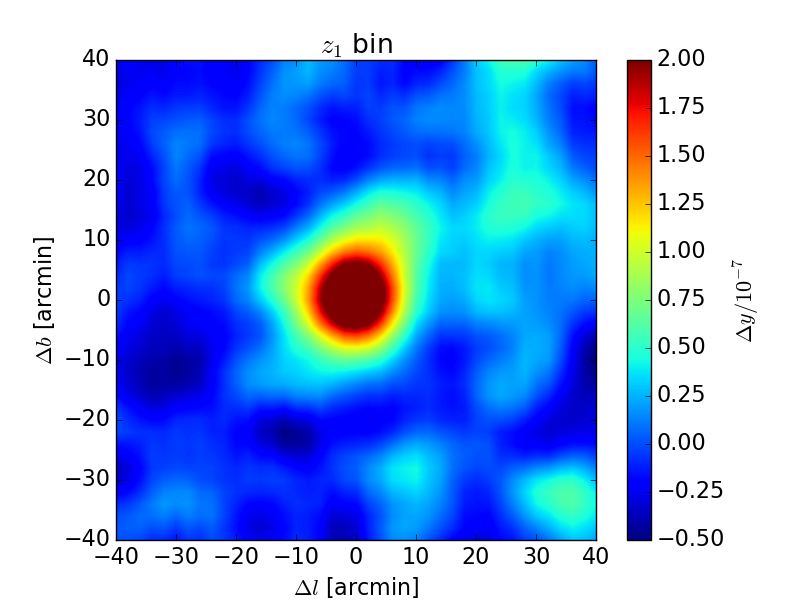}}
\resizebox{!}{!}{\includegraphics[scale=0.4]{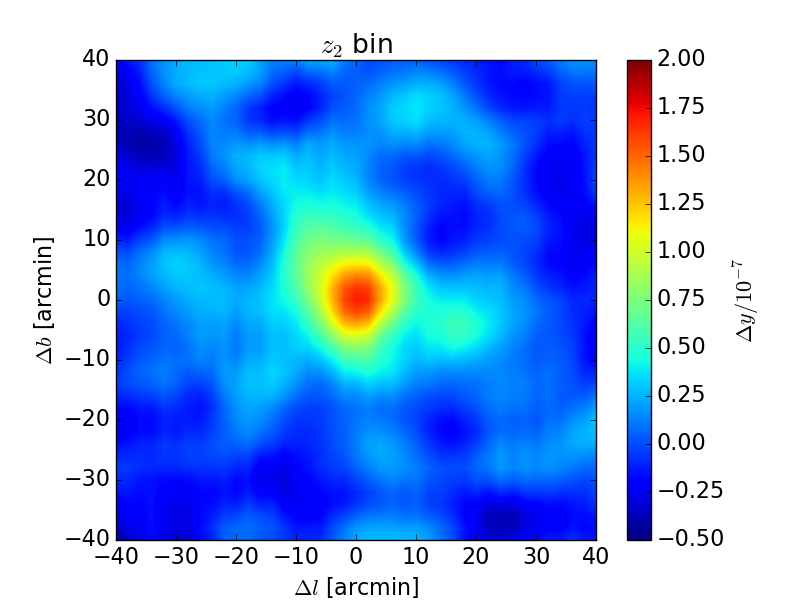}}
}
\centerline{
\resizebox{!}{!}{\includegraphics[scale=0.4]{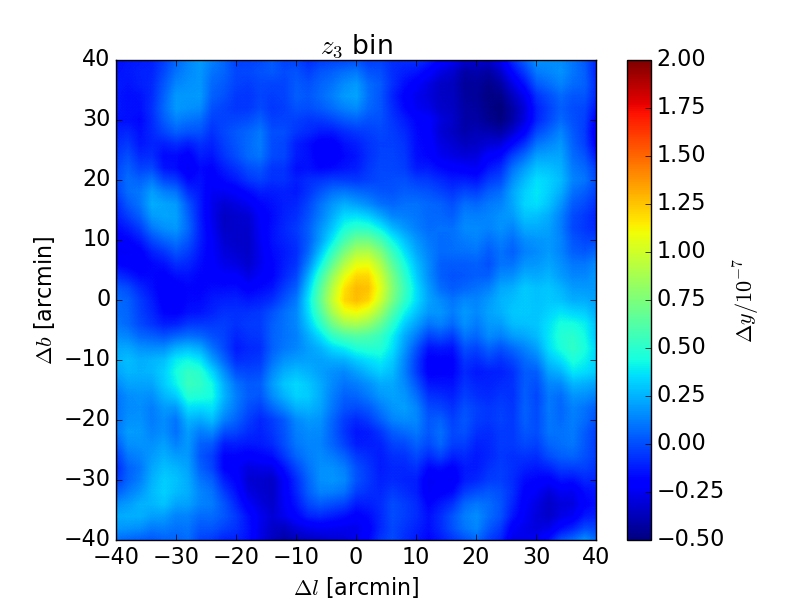}}
\resizebox{!}{!}{\includegraphics[scale=0.4]{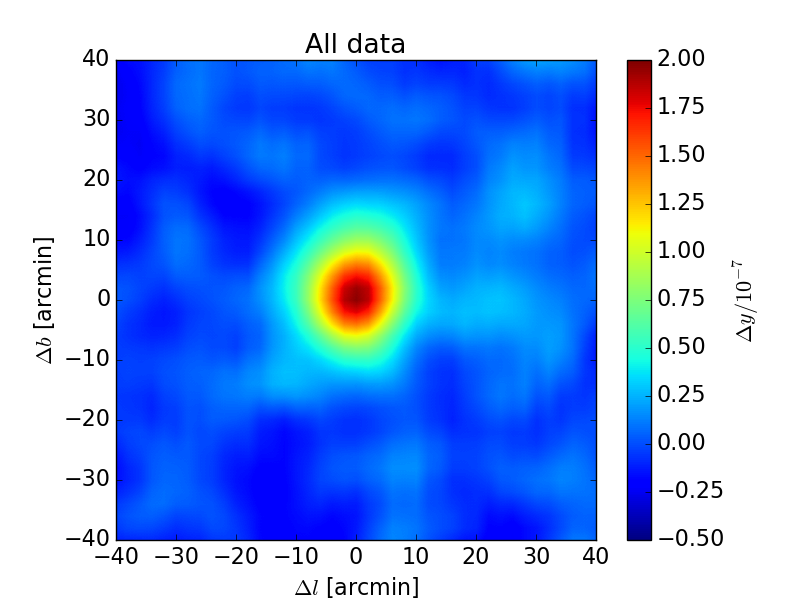}}
}
\caption{\label{fig:ymap} The stacked $y$ intensity maps for $z_1$ bin (upper left), $z_2$ bin (upper right), $z_3$ bin (bottom left), and the whole redshift range with all LRG data (bottom right). The number of selected LRGs stacked in each map is 11,660 with the same mass distribution as shown in Figure~\ref{fig:M_dis}. The mean tSZ intensity in the annular region between 30 and 40 arcmin has been subtracted as the local background.}
\end{figure*}

\section{Data}
\label{sec:data}
We adopt the same data sets used in \cite{Tanimura19b} in our analysis: the Luminous Red Galaxy catalogue from the Sloan Digital Sky Survey seventh data release \cite{Kazin2010} and the Planck Compton $y$ map \citep{Planck2016-XXII} from the 2015 data release \citep{Planck2016-I}. Each data is described briefly in this section.

\subsection{Luminous Red Galaxy catalogue}
\label{lrg}

LRGs are early-type, massive galaxies with little ongoing star formation, which are selected based on magnitude and colour cut. They are typically located in the centers of galaxy groups and clusters and have been used to detect and characterize the remnants of baryon acoustic oscillations (BAO) at low to intermediate redshift (\citealt{Eisenstein05, Kazin2010, Anderson14}). The SDSS DR7 LRGs are selected in \citet{Kazin2010} and cover $\sim$20\% of the sky with almost a flat distribution. The LRG catalogue provides 105,831 LRGs with galaxy positions, magnitudes and spectroscopic redshifts. Stellar masses of the LRGs are estimated in the New York University Value-Added catalogue (NYU-VAGC)\footnote{http://sdss.physics.nyu.edu/vagc/} using the K-correct software\footnote{http://howdy.physics.nyu.edu/index.php/Kcorrect} of \cite{Blanton2007} based on a stellar initial mass function of \cite{Chabrier2003} and stellar evolution synthesis models of \cite{Bruzual2003}. They are given in unit of $h^{-2}\msun$ and we adopt $h=0.6711$ \citep{Planck13-parameters}.

Since not all LRGs are located in the centers of galaxy clusters, some LRGs may reside in them as ``satellite'' galaxies. To minimize the satellite LRGs in our sample, we select locally (geometrically) most-massive LRGs (based on stellar mass) using a criterion that is analogous to that used in \cite{Planck2013IR-XI}: we remove a galaxy if a more massive galaxy resides within a tangential distance of 1.0 Mpc and within a radial velocity difference of $|c \Delta {\it z}| < 1000$ km s$^{-1}$. After this selection, 101,407 locally most-massive LRGs are left, that are likely to be ``central'' LRGs.  The details of the sample is described in \cite{Tanimura19b}.

\subsection{{\it Planck} $y$-map}

The {\it Planck} tSZ map from the {\it Planck} 2015 data release is provided in {\tt HEALPix} format \citep{Gorski2005} \footnote{http://healpix.sourceforge.net/} with a pixel resolution of $N_{\rm side}$ = 2048. Two types of algorithms, {\tt MILCA} \citep{Hurier13} and {\tt NILC} \citep{Remazeilles13}, are applied for the \Planck band maps to extract the tSZ signal. They are based on ILC (internal liear combination) techniques aiming to preserve an astrophysical component for which the electromagnetic spectrum is known \citep{Bennett03}. Our analysis is based on the {\tt MILCA} $y$-map, but we find consistent results with the {\tt NILC} $y$-map.

The 2015 data release also provides sky masks suitable for analyzing the $y$-maps. The mask covers a point-source mask and galactic masks that exclude 40\%, 50\%, 60\% and 70\%  of the sky. We combine the point source mask with the 40\% galactic mask, which excludes $\sim$50\% of the sky. The mask is used in our stacking process such that for a given LRG, masked pixels in the $y$-map near that LRG are not accumulated in the stacked image. Since the mask may bias the $y$ profile, we accept 74,681 LRGs, for which $\ge$80\% of the region within a 40 arcmin circle around each LRG is available.

\begin{figure*}
\centerline{
\resizebox{!}{!}{\includegraphics[scale=0.4]{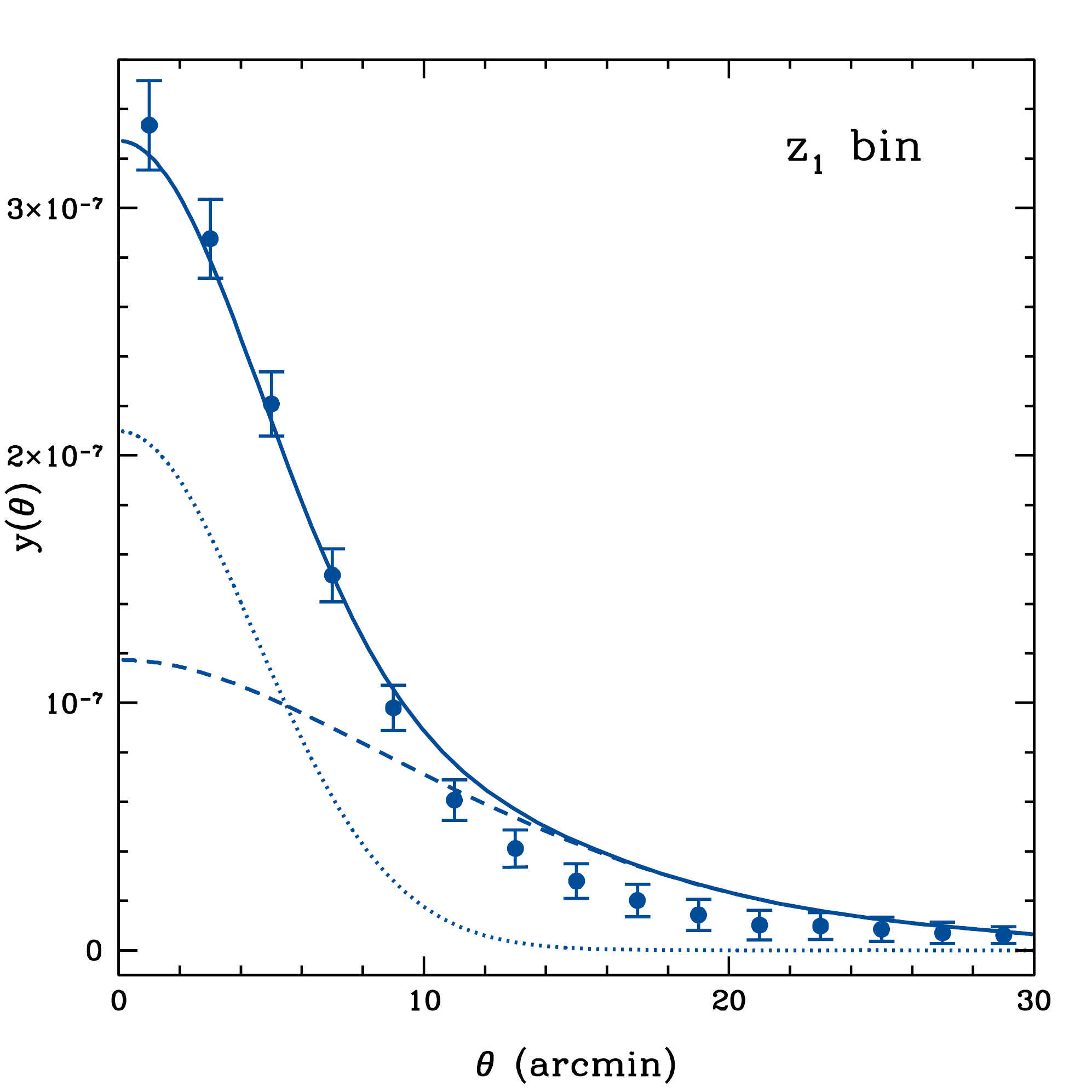}}
\resizebox{!}{!}{\includegraphics[scale=0.4]{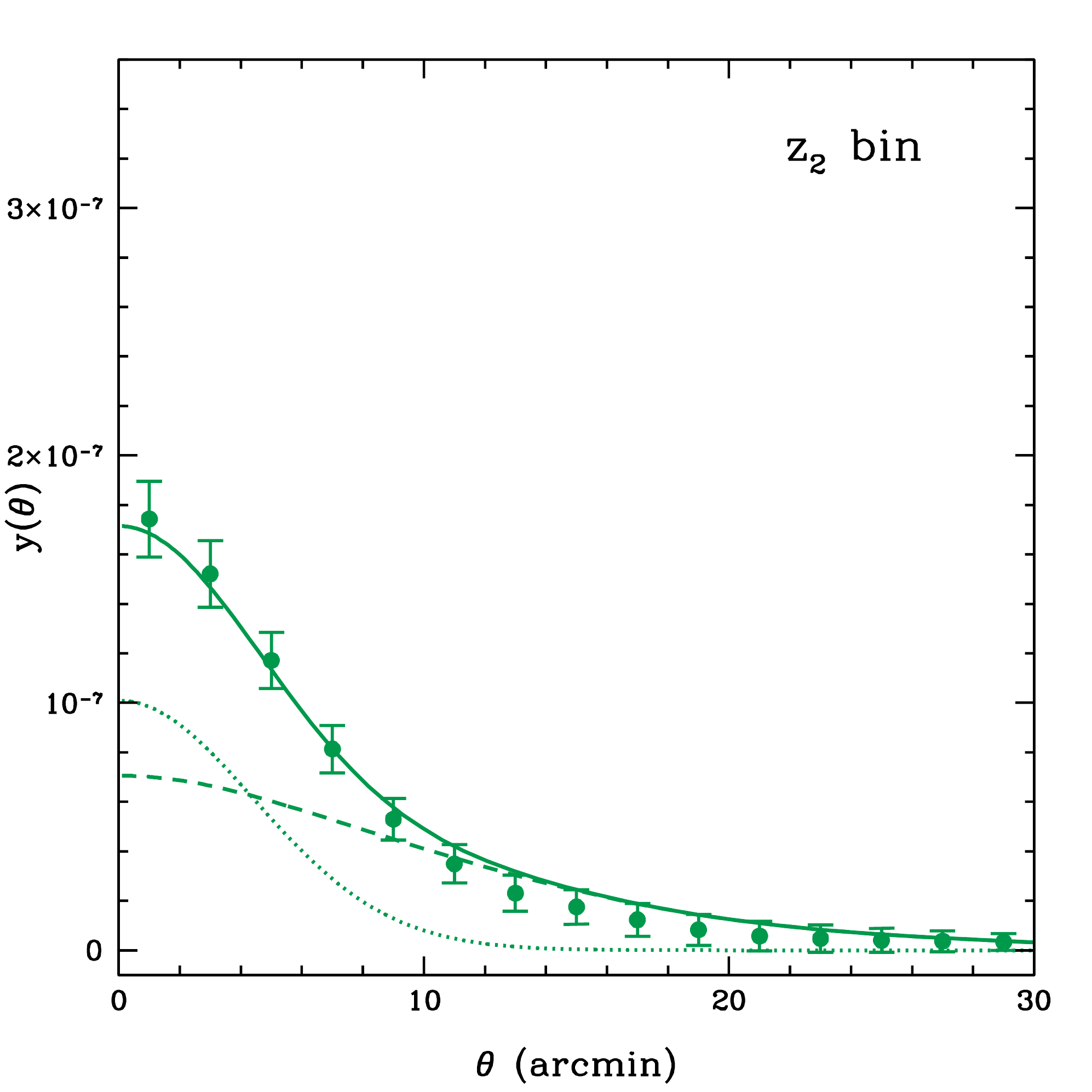}}
}
\centerline{
\resizebox{!}{!}{\includegraphics[scale=0.4]{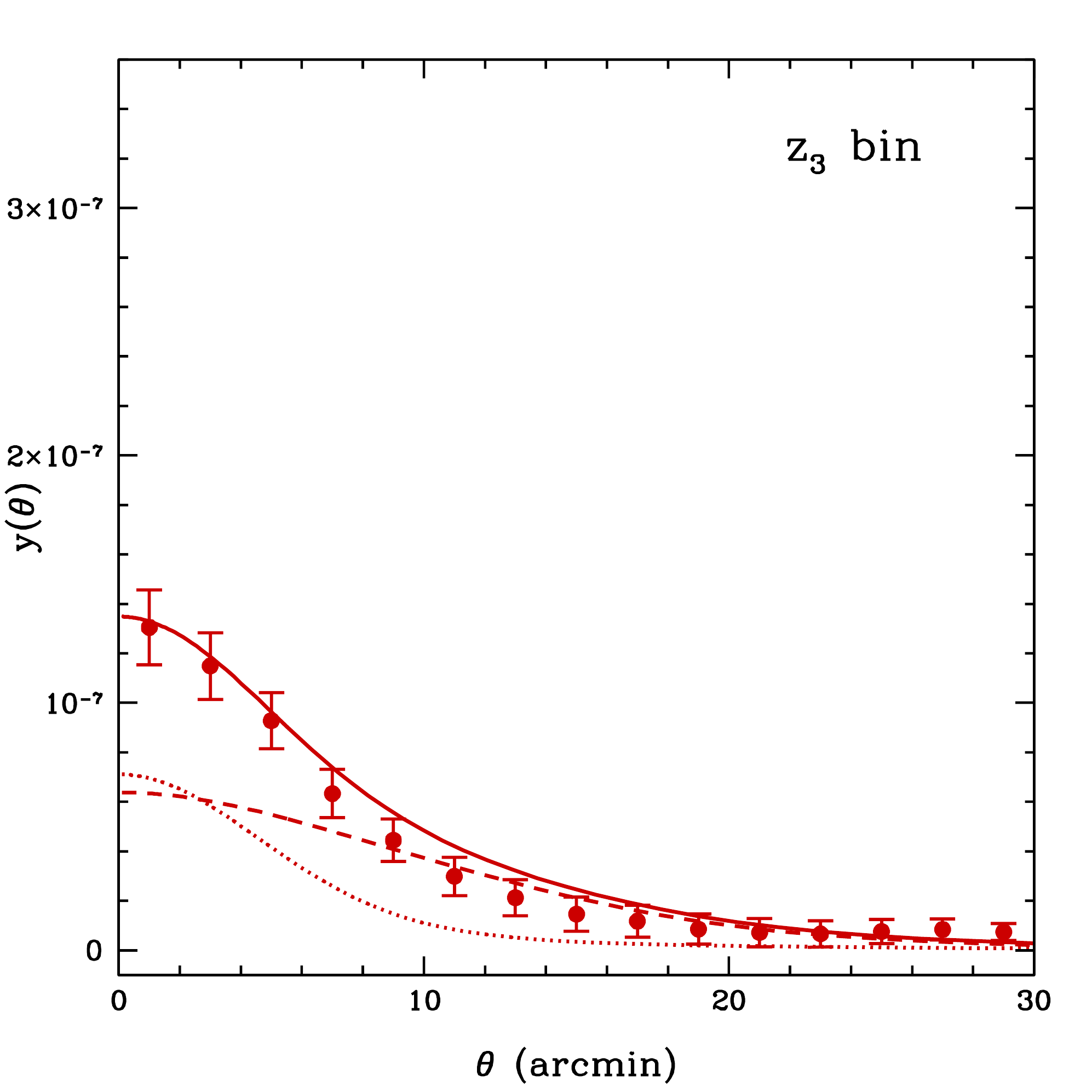}}
\resizebox{!}{!}{\includegraphics[scale=0.4]{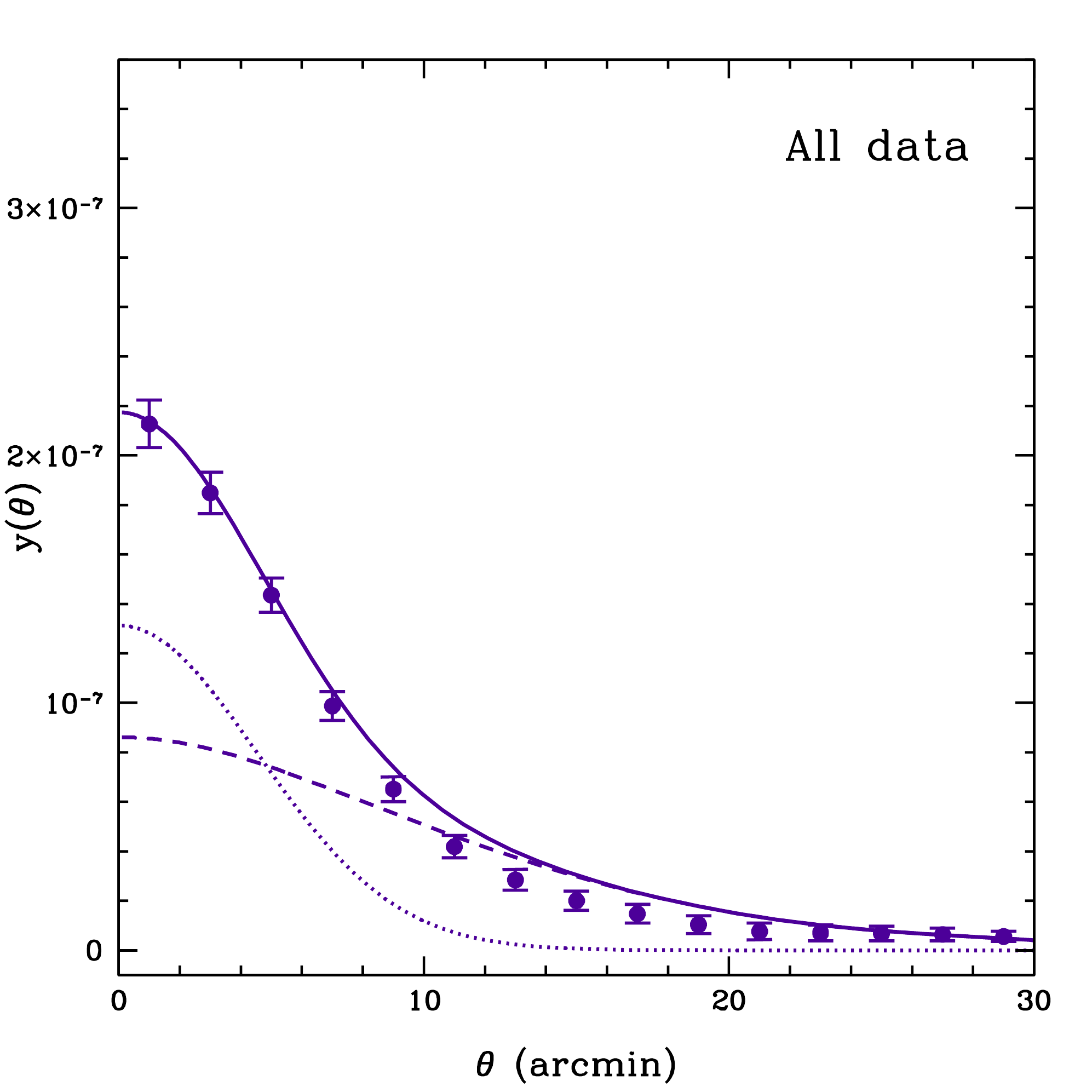}}
}
\caption{\label{fig:y_profile} The angular cross correlations of tSZ signal and galaxy cluster distribution in different redshift ranges. The data points are derived from the stacked intensity maps shown in Fig.~\ref{fig:ymap}. The error bars are simply derived from the diagonal elements of the covariance matrix in each case. The solid, dotted, and dashed curves are the best-fits of total, one-halo, and two-halo correlation functions, respectively. All of data points and curves are rescaled based on the average values of corresponding intensities between 30 and 40 arcmin.}
\end{figure*}

\subsection{Stacking $y$-map centered on LRG\lowercase{s}}
\label{subsec:stacked-map}

We describe our procedure for stacking the \Planck $y$-map at the positions of LRGs. We set a 2-dimensional angular coordinate system of $-40' < \Delta l < 40'$ and $-40' < \Delta b < 40'$, divided in 80 $\times$ 80 bins. For each LRG in the catalogue, we place it at the center of the coordinate and subtract its local background signal of the mean tSZ signal in the annular region between 30 and 40 arcmin. This procedure nulls a large scale mode of fluctuation in the $y$-map, not due to the target object. We repeat this procedure for our sample of LRGs, stack them and finally divide the stack by the number of our sample. 

In order to investigate a redshift evolution of tSZ signal (evolution of pressure profile), we divide our LRG sample into three redshift bins of sub-samples, $z_1$ bin (0.16<$z$<0.26), $z_2$ bin (0.26<$z$<0.36) and $z_3$ bin (0.36<$z$<0.47). These redshift bins are selected to have an equal interval in redshift to probe the redshift evolution of pressure profiles around $z \sim$ 0.2, 0.3 and 0.4. The number of LRGs in each redshift bin is 18,083 in $z_1$ bin, 29,586 in $z_2$ bin and 27,012 LRGs in $z_3$ bin. 

To study the redshift evolution of pressure profile, the effect due to differences of the mass distributions in different redshift bins (see Figure~\ref{fig:M_dis}) must be removed, and it needs to be probed by comparing the $y$ profiles with the ``same'' mass distribution. The original mass distributions of LRG halos at the three redshift bins are shown in Figure~\ref{fig:M_dis} in halo mass ($M_{500}$), for which halo masses are estimated by the stellar-to-halo mass relation in \cite{Wang16} obtained from gravitational lensing measurements. \citet{Tanimura19b} shows the $y$ profile of the LRGs can be described using this stellar-to-halo relation. We find that our constraint results are not quite sensitive to the stellar-to-halo relation. To obtain the same mass distributions in the three redshift bins, we find and select the minimum number of LRGs in a mass bin in the three $z$-bins (or the overlapped region of the three mass distributions, see Figure~\ref{fig:M_dis}), and adopt these LRGs as our sub-sample used in the following discussion. The derived mass distribution for the three $z$-bins is shown as hatched area in Figure~\ref{fig:M_dis} and it provides 11,660 LRGs for each redshift bin.

\begin{figure}
\centerline{
\resizebox{!}{!}{\includegraphics[scale=0.33]{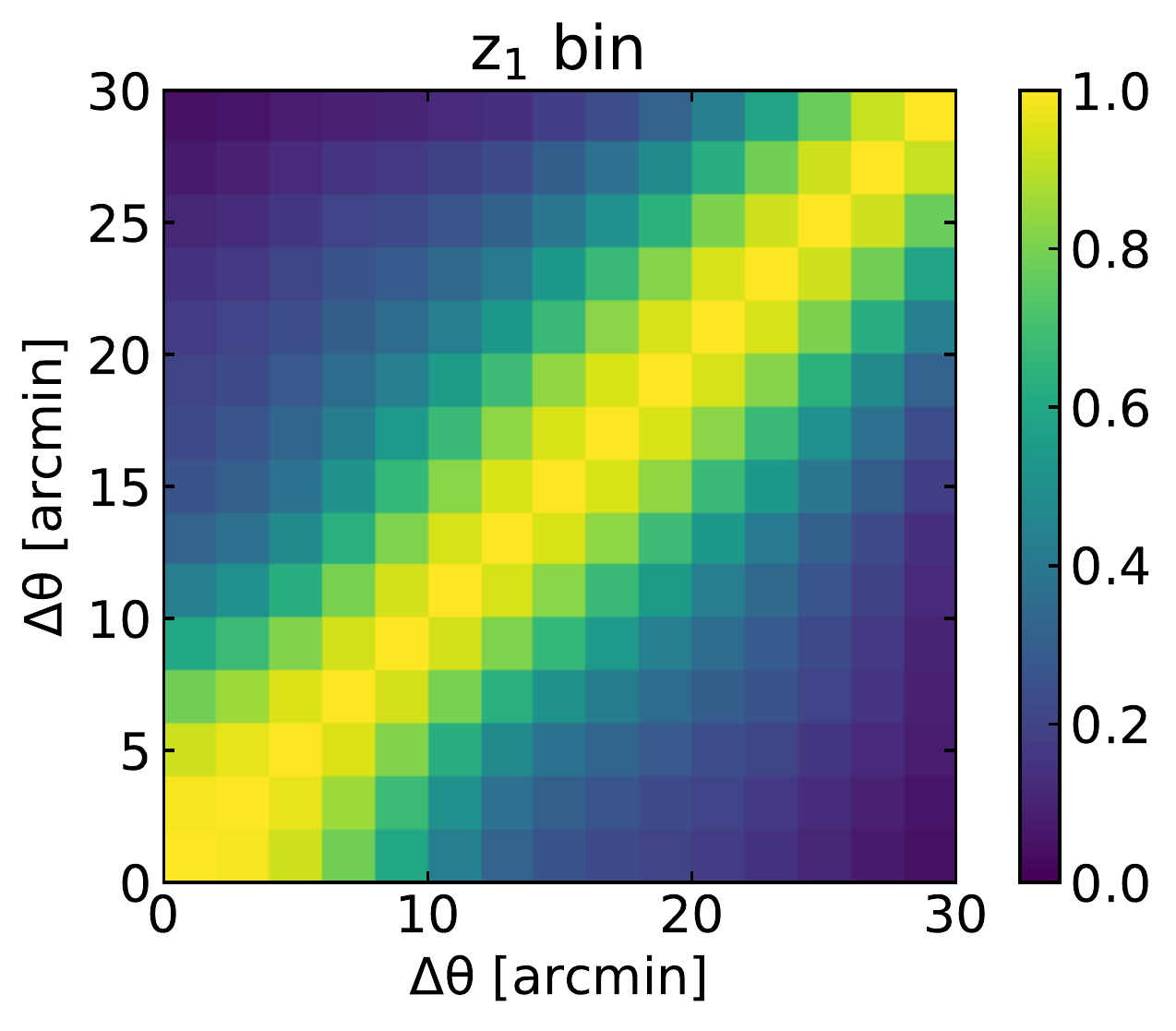}}
\resizebox{!}{!}{\includegraphics[scale=0.33]{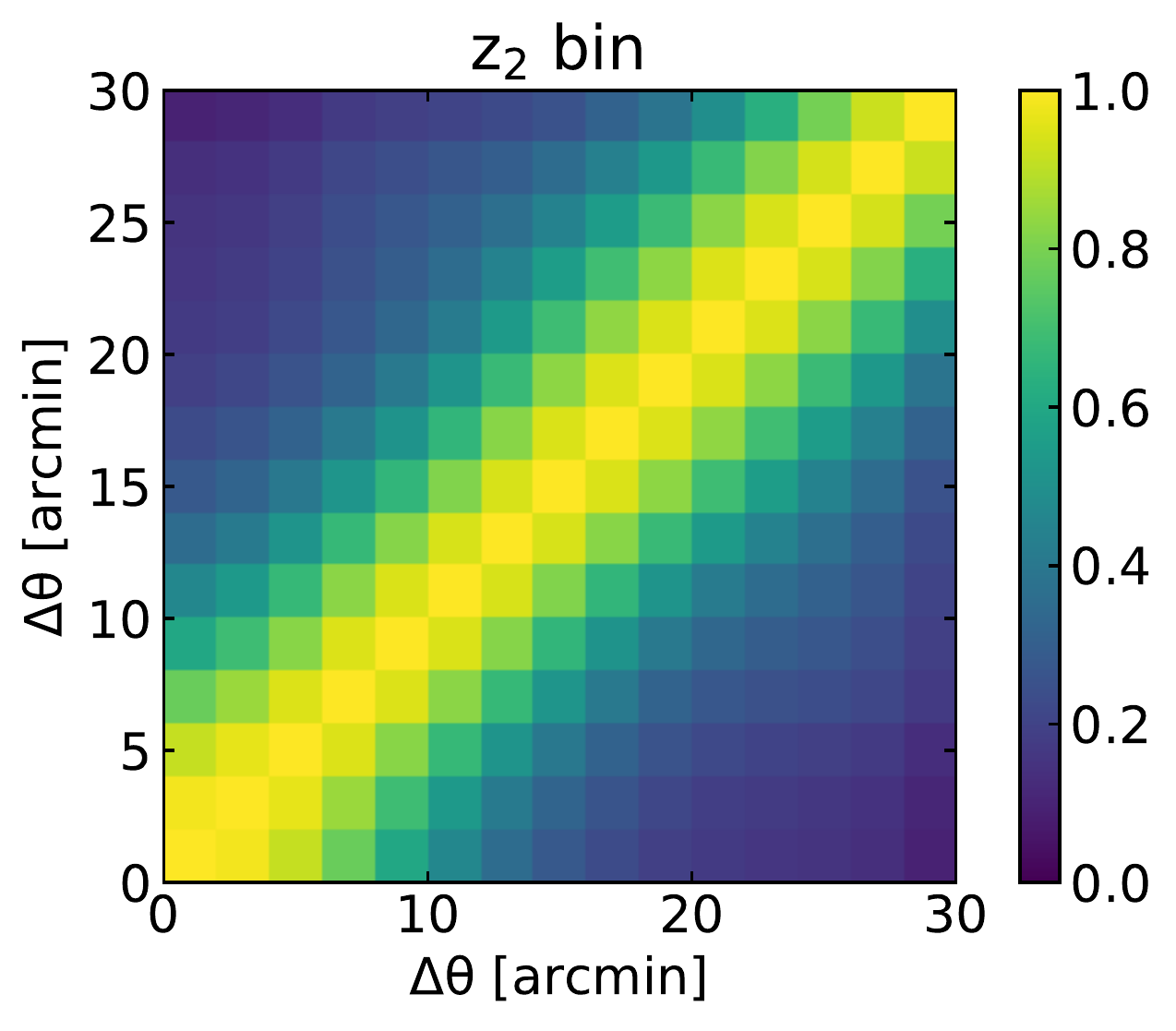}}
}
\centerline{
\resizebox{!}{!}{\includegraphics[scale=0.33]{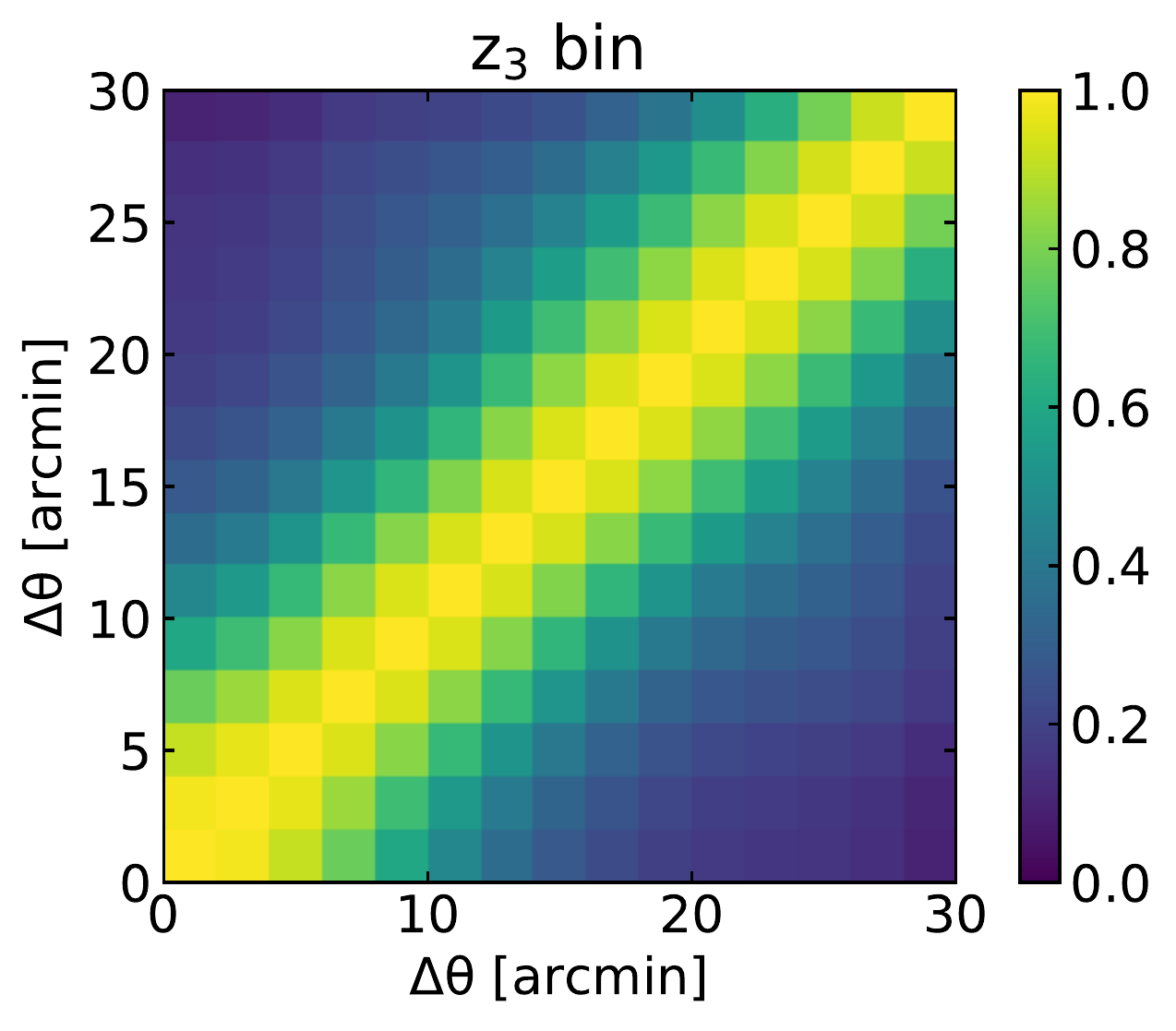}}
\resizebox{!}{!}{\includegraphics[scale=0.33]{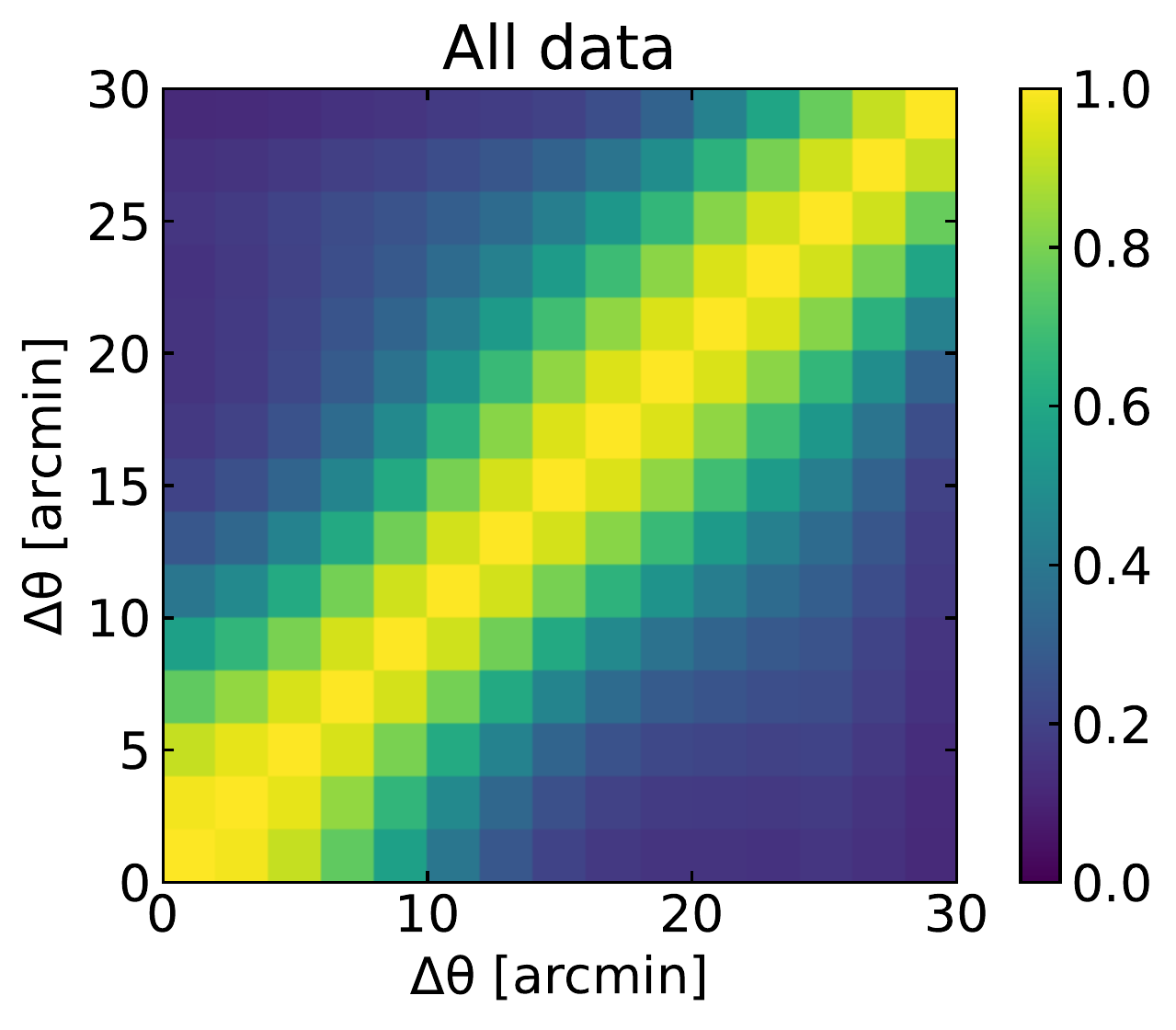}}
}
\caption{\label{fig:coeff} The correlation coefficient matrices between angular scales for the $y$ profiles $y(\theta)$ in different redshift ranges. The angular scale $\theta$ has been divided into 15 bins with $\Delta {\theta}=2$ arcmin.}
\end{figure}

Figure~\ref{fig:ymap} shows the average $y$-map stacked against 11,660 LRGs at $z_{1}$, $z_{2}$ and $z_{3}$ bin, respectively, with the same mass distribution. The corresponding $y$ profiles as a function of angular scale $\theta$ with their 1$\sigma$ statistical uncertainties are shown in Fig.~\ref{fig:y_profile}. The uncertainties are estimated by a bootstrap resampling by drawing a random sample of LRGs in a redshift bin. For example, in $z_{1}$ bin, 11,660 LRGs are resampled from the total of 18,083 LRGs with a replacement allowed. For the mass distribution to be unchanged, we replace a LRG with another LRG of the same mass. We repeat this process 1000 times and the bootstrapped data produce 1000 average $y$ profiles. The rms fluctuation of the profiles is shown as an uncertainty in Fig.~\ref{fig:y_profile}. We find that the average $y$ profiles have central peaks of $y = (3.3 \pm 0.2) \times 10^{-7}$ in $z_{1}$ bin, $y = (1.7 \pm 0.2) \times 10^{-7}$ in $z_{2}$ bin, $y = (1.3 \pm 0.2) \times 10^{-7}$ in $z_{3}$ bin, and $y =  (2.1 \pm 0.1) \times 10^{-7}$ for all data summed over the three redshift bins. As can be seen, we detect the tSZ signal out to $\sim$ 30 arcmin, that is well beyond the {\it Planck} beam of 10 arcmin. Note that there are correlations between different angular scales when deriving $y$ profiles. The correlation coefficient matrices in different redshift ranges are shown in Figure~\ref{fig:coeff}. Therefore, instead of using the independent errors shown in Figure~\ref{fig:y_profile}, we adopt the full covariance matrix to estimate the $\chi^2$ in the fitting process.

We also notice that, given the same angular scale $\theta$, $y(\theta)$ actually cover different physical scales of clusters at different redshifts. To investigate the stacked tSZ signal at the same cluster physical scales, and as a comparison, we also derive the stacked $y$ profiles as a function of $r/R_{500}$, where $r$ is the physical scale of radius. The corresponding $y(r/R_{500})$ results are shown in the Appendix. In principle, the results of constraints on the UPP parameters do not depend on whether physical or angular scale is chosen, as long as the same theoretical halo model is used to compare with the same data sets. Since it has great convenience for the $y(\theta)$ when convolving with the beam function (as shown in Sec. \ref{subsec:tSZ_theo}), and considering accuracy and time consumption in the computation process, we choose to use $y(\theta)$ in our following discussion.

\section{Stacked \lowercase{t}SZ profile modelling}
\label{sec:model}
\subsection{Compton $y$-parameter}

As shown in Eq.~(\ref{comptony}), the Compton $y$-parameter is an integral of the cluster's pressure profile along the line-of-sight. Therefore, for projected angle $\theta$ from the centre of the profile, the $y$-parameter can be calculated as~\citep{Komatsu11}
\begin{eqnarray}
y(\theta) = \frac{2\sigma_{\rm T}}{m_{\rm e}c^2}\int^{\sqrt{r^{2}_{\rm out}-\theta^{2}D^{2}_{\rm A}}}_{0} P_{\rm e}\left(\sqrt{l^{2}+\theta^{2}D^{2}_{\rm A}} \right) \der l,
\label{eq:yt}
\end{eqnarray} 
where $D_{\rm A}(z)$ is the angular diameter distance to the LRG sample. The reason that there is a factor of $2$ in front of the integral and the lower limit is zero is because of the spherical symmetry. The upper limits of the integral is $\sqrt{r^2_{\rm out}-\theta^2D_{\rm A}^2}$, where $r_{\rm out}$ is a truncated scale and we always have $r_{\rm out}\gg \theta D_{\rm A}$ \citep{Komatsu11}. 
The electron UPP $P_{\rm e}$ can be written as
\begin{eqnarray} 
P_{\rm e}(r) = P'_{500}\,\mathbb{P}(x),
\label{eq:Pe}
\end{eqnarray}
where $x\equiv r/R_{500}$, and $R_{500}$ is the radius within which the average density is $500$ times higher than the critical density of the Universe. The $P'_{500}=P_{500}F_{500}$, where $P_{500}$ is a characteristic pressure, based on the standard self-similar model for the variation of cluster mass \citep{Arnaud10}
\begin{eqnarray} \label{eq:P500}
P_{500} &=& 1.65\times 10^{-3} E(z)^{8/3} \nonumber \\
              &\times& \left[ \frac{M_{500}}{3\times10^{14}\,h_{70}^{-1}\msun}\right]^{2/3} h_{70}^2 \ \rm{keV\, cm^{-3}},
\end{eqnarray}
where $h_{70}=h/0.7$ and $E(z)=H(z)/H_0$. The $F_{500}$ is a correction factor of $P_{500}$, reflecting the deviation of the standard self-similar scaling, given by~\citet{Arnaud10} and~\citet{Planck13}
\begin{equation}
F_{500} = \left[ \frac{M_{500}}{3\times10^{14}\,h_{70}^{-1}\msun} \right]^{0.12}.
\end{equation}
The $\mathbb{P}(x)$ in Eq.~(\ref{eq:Pe}) is the scaled pressure profile, normalized by $P_{500}$. By adopting the generalized NFW profile \citep{Navarro97,Nagai07}, it can be characterized as
\begin{equation}
\mathbb{P}(x) = \frac{P_0}{(c_{500}x)^{\gamma}\left[ 1+(c_{500}x)^\alpha \right]^{(\beta-\gamma)/\alpha}}, \label{eq:ddP}
\end{equation}
where $P_0$ is a normalization factor, $c_{500}$ is the concentration factor at $R_{500}$, and $\gamma$, $\alpha$, and $\beta$ are the slopes for the central, intermediate, and outer regions of cluster, respectively. Following \cite{Planck13}, we fix $\gamma=0.31$, and fit the rest four parameters in the fitting process.

\subsection{Theoretical stacked tSZ profile}
\label{subsec:tSZ_theo}
The observed angular cross correlation function of tSZ signal and galaxy cluster distribution, i.e. stacked tSZ profile, ${y}_{\rm cross}(\theta)$ can be predicted by expanding it into Legendre polynomials considering beam function of an experiment \citep[see e.g.][]{Komatsu99}
\begin{eqnarray} 
{y}_{\rm th}^{\rm cross}(\theta) = \frac{1}{4\pi} \sum_{\ell} (2\ell+1) C_{\ell}^{yc} \,P_{\ell}({\rm cos}\,\theta)\, B_{\ell},
\label{eq:y_th}
\end{eqnarray}
where $P_{\ell}(x)$ are the Legendre polynomials, $C_{\ell}^{yc}$ is the angular cross power spectrum of tSZ and cluster distribution, and the beam function $B_{\ell}={\rm exp}(-\ell^2\sigma_{\rm b}^2/2)$. Here $\sigma_{\rm b}=\theta_{\rm FWHM}/\sqrt{8\,{\rm ln}2}$, where we take $\theta_{\rm FWHM}=10$ arcmin as the \Planck beam size. By adopting the flat-sky and Limber approximations, the $C_{\ell}^{yc}$ can be expressed as \citep{Cole88,Komatsu99,Fang12}
\begin{eqnarray} 
C_{\ell}^{yc} = C_{\ell}^{yc, \rm 1h} + C_{\ell}^{yc, \rm 2h},
\label{eq:Cl_yc}
\end{eqnarray}
where $C_{\ell}^{yc, \rm 1h}$ and $C_{\ell}^{yc, \rm 2h}$ are the one-halo and two-halo terms, respectively. 

Theoretically, following \cite{Fang12}, in given redshift and cluster mass ranges, the one- and two-halo terms take the forms as
\begin{eqnarray} 
C_{\ell}^{yc, \rm 1h} = \frac{1}{\bar{n}_{\rm2D}}\int_{z_{\rm l}}^{z_{\rm u}} {\rm d}z \frac{c\, \chi^2}{H(z)} \int_{M_{\rm l}}^{M_{\rm u}} {\rm d}M \frac{\der n}{\der M}\, y_{\ell}(M,z),
\label{eq:Cl_1h}
\end{eqnarray}
\begin{eqnarray} 
C_{\ell}^{yc, \rm 2h} = \frac{1}{\bar{n}_{\rm2D}} \int_{z_{\rm l}}^{z_{\rm u}} {\rm d}z \frac{c\, \chi^2}{H(z)} P_{\rm m}\left( k, z \right) W^c(z)W^y_{\ell}(z). 
\label{eq:Cl_2h}
\end{eqnarray}
Here $z_{\rm l}$ and $z_{\rm u}$, $M_{\rm l}$ and $M_{\rm u}$ are the lower and upper bounds of redshift and mass ranges, respectively. $\chi$ is the comoving distance, $b(M,z)$ and $\der n/ \der M$ are the halo bias and mass function \citep{Sheth99}, and $P_{\rm m}(k,z)$ is the linear matter power spectrum, where $k=(\ell+1/2)/\chi$. The $\bar{n}_{\rm2D}$ is the two-dimensional angular number density of galaxy cluster in given redshift and cluster mass ranges, that can be estimated by~\citep{Fang12}
\begin{equation}
\bar{n}_{\rm2D} = \int_{z_{\rm l}}^{z_{\rm u}} {\rm d}z \frac{c \chi^2}{H(z)} \int_{M_{\rm l}}^{M_{\rm u}} {\rm d}M \frac{{\rm d}n}{{\rm d}M}(M,z).
\end{equation}
The functions $W^c(z)$ and $W^y_{\ell}(z)$ are defined as
\begin{eqnarray}
W^c(z) & \equiv & \int_{M_{\rm l}}^{M_{\rm u}} {\rm d}M \frac{{\rm d}n}{{\rm d}M}(M,z)\, b(M,z), \nonumber \\
W^y_{\ell}(z) & \equiv & \int {\rm d}M \frac{{\rm d}n}{{\rm d}M}(M,z)\, b(M,z)\, y_{\ell}(M,z), \label{eq:Wy}
\end{eqnarray}
where the second integral should cover all possible halo masses as it is the influence from all other correlated halos. The $y_{\ell}(M,z)$ is the Fourier transform of the Compton $y$-parameter, written as
\begin{eqnarray} 
y_{\ell}(M,z) = \frac{a}{\chi^2}\left( \frac{\sigma_{\rm T}}{m_{\rm e}c^2} \right) u_{\rm p} \left( k=\left.\frac{\ell+1/2}{\chi} \right| M,z \right),
\label{eq:yl}
\end{eqnarray}
and
\begin{eqnarray} 
u_{\rm p} = \int {\rm d}r \ 4\pi r^2  \frac{{\rm sin}(kr)}{kr} P_{\rm e}(r|M,z).
\label{eq:up}
\end{eqnarray}
Note that the $P_{\rm e}(r|M,z)$ here is based on virial mass $M_{\rm vir}$ instead of $M_{500}$ used in Eq.~(\ref{eq:Pe}), so we need to calculate this $P_{\rm e}$ by converting $M_{\rm vir}$ to $M_{500}$ \citep[e.g. see appendix B in][]{Planck17}. Then we can estimate the angular cross power spectrum $C_{\ell}^{yc}$ theoretically using Eqs.~(\ref{eq:Cl_1h})-(\ref{eq:up}) with the help of halo model \citep{Cooray02}.

\begin{table*}
\caption{The best-fits and 1$\sigma$ errors of the free parameters in the UPP model. The parameter $\gamma$ is fixed to be 0.31 in the fitting process \citep{Planck13}. The results from~\citet{Arnaud10} (A10) and \citet{Planck13} (Planck13) are also shown at the bottom as comparison.}
\label{tab:fit}
\vspace{1mm}
\begin{center}
\begin{tabular}{c|c|c|c|c|c|c|c}
\hline  \hline
Case & $P_0$ & $c_{500}$ & $\alpha$ & $\beta$ & $\gamma$ & $\eta$ & $\rm \chi^2_{\rm red}$ \\
\hline
$z_1$ bin & $3.35^{+8.18}_{-1.35}$ & $1.45^{+0.56}_{-0.41}$ &  $4.31^{+0.66}_{-2.88}$ & $5.14^{+3.00}_{-0.86}$  & 0.31 & - &  1.65\\
\hline
$z_2$ bin & $10.46^{+1.54}_{-6.85}$ &$1.88^{+0.53}_{-0.55}$  & $4.03^{+0.96}_{-2.79}$  & $4.39^{+3.45}_{-0.79}$  & 0.31 & - & 0.45\\
\hline
$z_3$ bin & $6.15^{+5.25}_{-3.67}$ & $2.23^{+2.42}_{-0.80}$ & $1.24^{+3.68}_{-0.50}$  & $3.00^{+0.43}_{-0.30}$  & 0.31 & - &  0.52\\
\hline
All data & $2.18^{+9.02}_{-1.98}$ & $1.05^{+1.27}_{-0.47}$ & $1.52^{+1.47}_{-0.58}$ &  $3.91^{+0.87}_{-0.44}$ & 0.31 & - &  1.88\\
\hline
3 bins & $3.04^{+15.53}_{-1.93}$ & $2.12^{+0.82}_{-1.01}$ & $5.80^{+3.89}_{-4.46}$ & $4.83^{+2.20}_{-0.78}$ & 0.31 & - & 1.61\\
\hline
3 bins with $\eta$ & $2.99^{+3.44}_{-1.57}$ & $1.16^{+0.79}_{-0.29}$ & $2.66^{+1.67}_{-0.97}$ & $5.48^{+2.39}_{-1.38}$ & 0.31 & $-3.11^{+1.09}_{-1.13}$ & 1.39 \\
\hline
A10 & 8.403\,$h_{70}^{-3/2}$ & 1.177 & 1.0510 & 5.4905 & 0.3081 & - & - \\
\hline
Planck13 & 6.41 & 1.81 & 1.33 & 4.13 & 0.31& - & - \\
\hline
\end{tabular}
\end{center}
\end{table*}

\subsection{Predicted stacked tSZ profile}
\subsubsection{Separating into mass and redshift bins}
Since we have large number of observed LRGs in SDSS DR7 (in total $74,681$, and $18,083$, $29,586$, and $27,012$ samples in $z_{1}$, $z_{2}$ and $z_{3}$ redshift bins respectively), it is quite time-consuming to calculate $\bar{y}(\theta)$ in a redshift range. In order to speed up the calculation, we further
divide each redshift bin ($z_{1}$, $z_{2}$ and $z_{3}$) into 4 sub-bins, and divide the total mass range of clusters into $10$ different mass bins. We compute the average redshift $\bar{z}_i$ for a redshift bin $i$, and average cluster mass $\overline{M}^j_{500}$ and radius $\bar{R}^j_{500}$ for a mass bin $j$, and estimate the $P_{\rm e, {\it i}}^{(j)}(r| \overline{M}^j_{500}, \bar{R}^j_{500})$ and $y_{i}^{(j)}(\theta| \overline{M}^j_{500}, \bar{R}^j_{500})$ using Eq.~(\ref{eq:Pe}) and Eq.~(\ref{comptony}) for redshift bin $i$ and mass bin $j$ (in below, we shorten it as ``the $ij$-th bin''). We also count the number of samples $n^{(j)}_{i}$ within the $ij$-th bin. Then the mean Compton $y$-parameter can be calculated as
\begin{eqnarray}
\bar{y}(\theta) = \frac{1}{N_{\rm c}} \sum_i^{N_z} \sum_j^{N_{M}} f_M^{(j)} n^{(j)}_{i}\, y^{(j)}_{i}(\theta). \label{eq:y-theta2}
\end{eqnarray} 
Here $N_z =4$ is the number of redshift sub-bins for either $z_{1}$, $z_{2}$ or $z_{3}$ bin ($N_{z}=12$ for the whole sample). $N_{M}=10$ is the number of cluster mass bins. $f_M^{(j)}$ is the number fraction of the selected sample in a mass bin. $N_{\rm c}$ is the total number of clusters, given by
\begin{eqnarray}
N_{\rm c}=\sum_i^{N_z} \sum_j^{N_{M}} n^{(j)}_{i}
\end{eqnarray}

Using such a binning strategy, the computation speed becomes faster since we don't need to calculate the $\mathcal{O}(10^{4})$ number of $y(\theta)$ profile individually. To verify the accuracy of such approach, we compute the $\bar{y}(\theta)$ for such binning strategy and the individual sum up, and the difference is only at $\sim$1\% at most in different redshift ranges. Hence, we will use this scheme to compute the one-halo term as follows (see Eq.~(\ref{eq:Cl_1h_p})).

\subsubsection{One-halo term}
Since we have the information of LRG redshift $z$, $M_{500}$, and $R_{500}$, we can directly predict the one-halo term for each cluster, instead of using the halo mass function $\der n/\der M$ in Eq.~(\ref{eq:Cl_1h}). 

We first count the number of LRGs for each sub-bin of mass and redshift, i.e. counting $n^{(j)}_{i}$, and then calculate the $y^{(j)}_{i}(\theta)$ profile for each sub-bin. Then we use Eq.~(\ref{eq:y-theta2}) to calculate the average of the profile within each redshift bin ($z_{1}$, $z_{2}$, $z_{3}$ or the whole redshift range). This profile is the unconvolved, averaged profile for a certain redshift bin. Because for real data, it is the original Compton-$y$ profile convolved with {\it Planck} beam, so we need to calculate the convolved averaged profile with {\it Planck} beam function. For this reason, we first Fourier-transform the averaged profile into the $\ell$-space of one-halo term $C_{\ell}^{yc, \rm 1h}$, i.e.
\begin{eqnarray} 
C_{\ell}^{yc, \rm 1h} = 2\pi \int_{-1}^1 P_{\ell}({\rm cos}\theta)\, \bar{y}(\theta)\, {\rm d}\, {\rm cos}\theta,
\label{eq:Cl_1h_p}
\end{eqnarray}
where $P_{\ell}$ are the Legendre polynomials. Then we calculate the following two-halo term by using halo model.

\subsubsection{Two-halo term}
Similarly, making use of the same binning method, we can also calculate the two-halo term $C_{\ell}^{yc, \rm 2h}$, and  we have
\begin{eqnarray} 
C_{\ell}^{yc, \rm 2h} = \frac{1}{N_{\rm c}}  \sum_i^{N_z} \sum_j^{N_M} f_M^{(j)} n_{i}^{(j)} C_{\ell}^{yc, \rm 2h}(\overline{M}_j, \bar{z}_i),
\label{eq:Cl_2h_p}
\end{eqnarray}
where $C_{\ell}^{yc, \rm 2h}(\overline{M}_j, \bar{z}_i)$ is the two-halo term in the  mass bin $j$ and redshift bin $i$. Note that $\overline{M}_j$ is an average virial mass, and can be obtained by $\overline{M}^j_{500}$.
The $C_{\ell}^{yc, \rm 2h}$ can be computed using Eq.~(\ref{eq:Cl_2h}) considering binning. In each redshift and mass bin, we notice that $\int^{z_{u}}_{z_{l}}\der z \longrightarrow \Delta z_{i}$ and $\int^{M_{u}}_{M_{l}}\der M \longrightarrow \Delta M_{j}$,
where $\Delta z_{i}$ and $\Delta M_{j}$ are the bin widths of the redshift bin $i$ and mass bin $j$, respectively. Therefore we can simplify the equation as
\begin{eqnarray}
C_{\ell}^{yc, \rm 2h}(\overline{M}_j, \bar{z}_i) = b(\overline{M}_j,\bar{z}_i) P_{\rm m}\left(k=\frac{\ell+1/2}{\chi(\bar{z}_i)}, \bar{z}_i \right)W^y_{\ell}(\bar{z}_i).
\end{eqnarray}

Therefore, the predicated stacked tSZ profile can be calculated by adding together Eq.~(\ref{eq:Cl_2h_p}) with Eq.~(\ref{eq:Cl_1h_p}) (i.e. Eq.~(\ref{eq:Cl_yc})), and then Fourier transform it back to real space with multiplication of {\it Planck} beam function through Eq.~(\ref{eq:y_th}). Besides, in Sect.~\ref{subsec:stacked-map}, we showed that in order to subtract background we subtract the mean value of the measured profile in the ring of $30$--$40$ arcmin. Therefore, in order to make accurate comparison, we also need to subtract the mean value for theoretical profile, i.e.
\begin{eqnarray}
\tilde{y}_{\rm th}^{\rm cross}(\theta) = {y}_{\rm th}^{\rm cross}(\theta) - \bar{y}_{\rm th}^{\rm cross}|_{\rm BG},
\end{eqnarray}
where $\bar{y}_{\rm th}^{\rm cross}|_{\rm BG}$ is the mean value of ${y}_{\rm th}^{\rm cross}(\theta)$ in the range of $30$--$40$ arcmin as the background value. We calculate $\tilde{y}_{\rm th}^{\rm cross}(\theta)$ in the whole redshift range and three main redshift bins respectively for sampling of the UPP parameters $(P_{0},c_{500},\alpha,\beta)$, and fit the corresponding datasets to derive the probability distribution functions (PDFs) of the these parameters.

\section{Fitting method}
\label{sec:analysis}
In order to fit the measured stacked tSZ profile $y_{\rm obs}^{\rm cross}(\theta)$ and extract the parameter values of the UPP $P_{\rm e}(r)$, we adopt the $\chi^2$ statistic method, given by
\begin{eqnarray}
\chi^2 &=&\sum_{m,n}^{N_{\theta}}\ [y_{\rm obs}^{\rm cross}(\theta_m)-\tilde{y}_{\rm th}^{\rm cross}(\theta_m)]\ (C_{\theta \theta'})_{mn}^{-1}  \nonumber\\
            &\times& [y_{\rm obs}^{\rm cross}(\theta_n)-\tilde{y}_{\rm th}^{\rm cross}(\theta_n)],
\end{eqnarray}
where $N_{\theta}$ is the number of data at different angular scales, and $C_{\theta \theta'}$ is the covariance matrix. Then we can calculate the likelihood function as $\mathcal{L}\sim {\rm exp}(-\chi^2/2)$.

We employ the Markov Chain Monte Carlo (MCMC) method to perform constraints on the free parameters in the UPP. The Metropolis-Hastings algorithm is adopted to determine the accepting probability of the new chain points \citep{Metropolis53,Hastings70}. The proposal density matrix is obtained by a Gaussian sampler with adaptive step size \citep{Doran04}. We assume uniform flat priors for all the free parameters, and their ranges are set to be $P_0\in(0,20)$, $c_{500}\in(0,10)$, $\alpha\in(0,10)$, and $\beta\in(0,10)$. We also add a free parameter $\eta$ on the power-law index of $E(z)$ in Eq.~(\ref{eq:P500}), when simultaneously fitting the three redshift bins together. This parameter can adjust the redshift-dependence of the electron pressure profile, and we set $\eta\in(-10,10)$. We run fifteen parallel chains for each case we explore, and obtain $10^5$ chain points for one chain after it reaches the convergence criterion \citep{Gelman92}. After performing the burn-in and thinning the chains, we merge all chains together and obtain about 10,000 chain points to illustrate one-dimensional (1-D) and two-dimensional (2-D) PDFs of the free parameters.

\begin{figure*}
\centerline{
\resizebox{!}{!}{\includegraphics[scale=0.41]{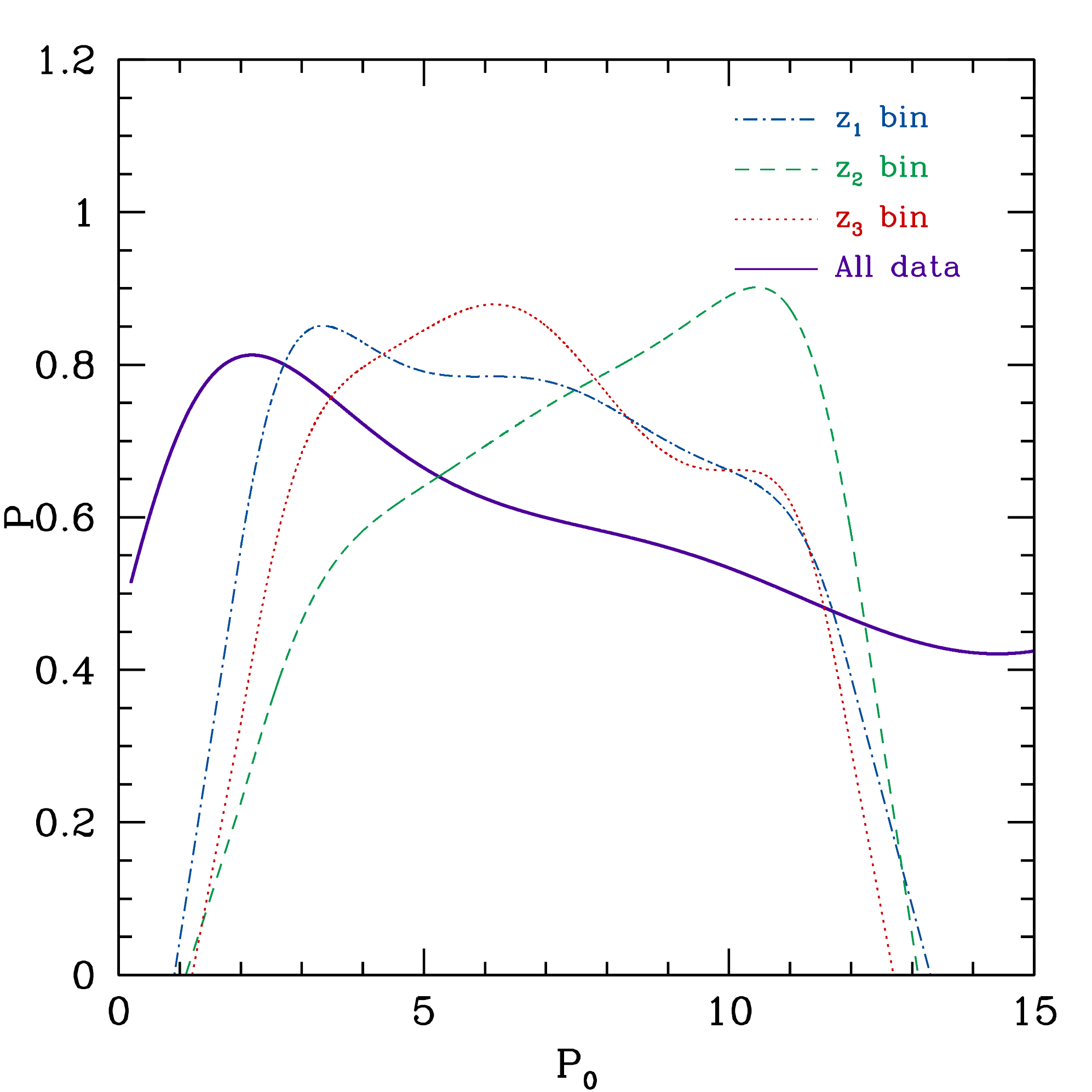}}
\resizebox{!}{!}{\includegraphics[scale=0.41]{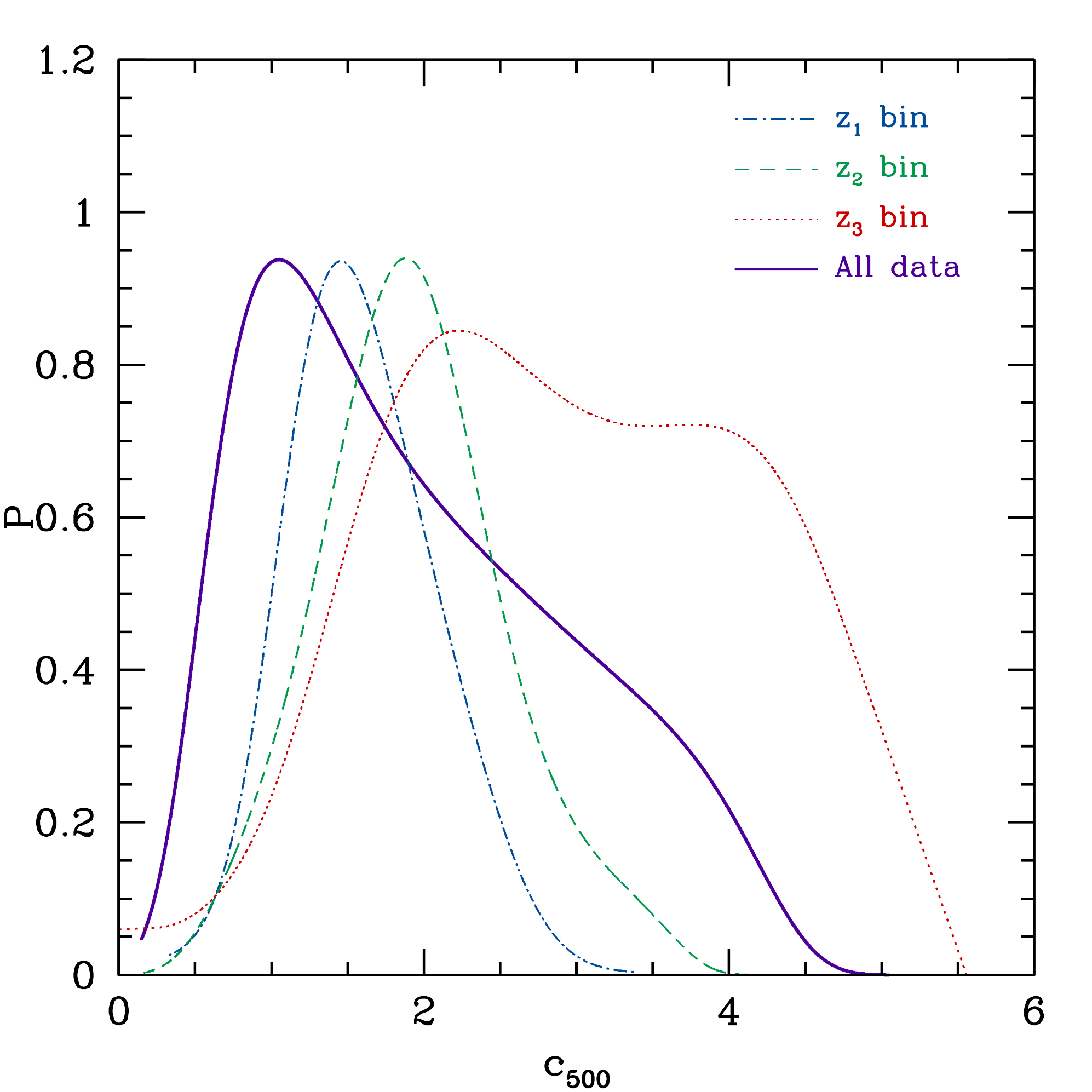}}
}
\centerline{
\resizebox{!}{!}{\includegraphics[scale=0.41]{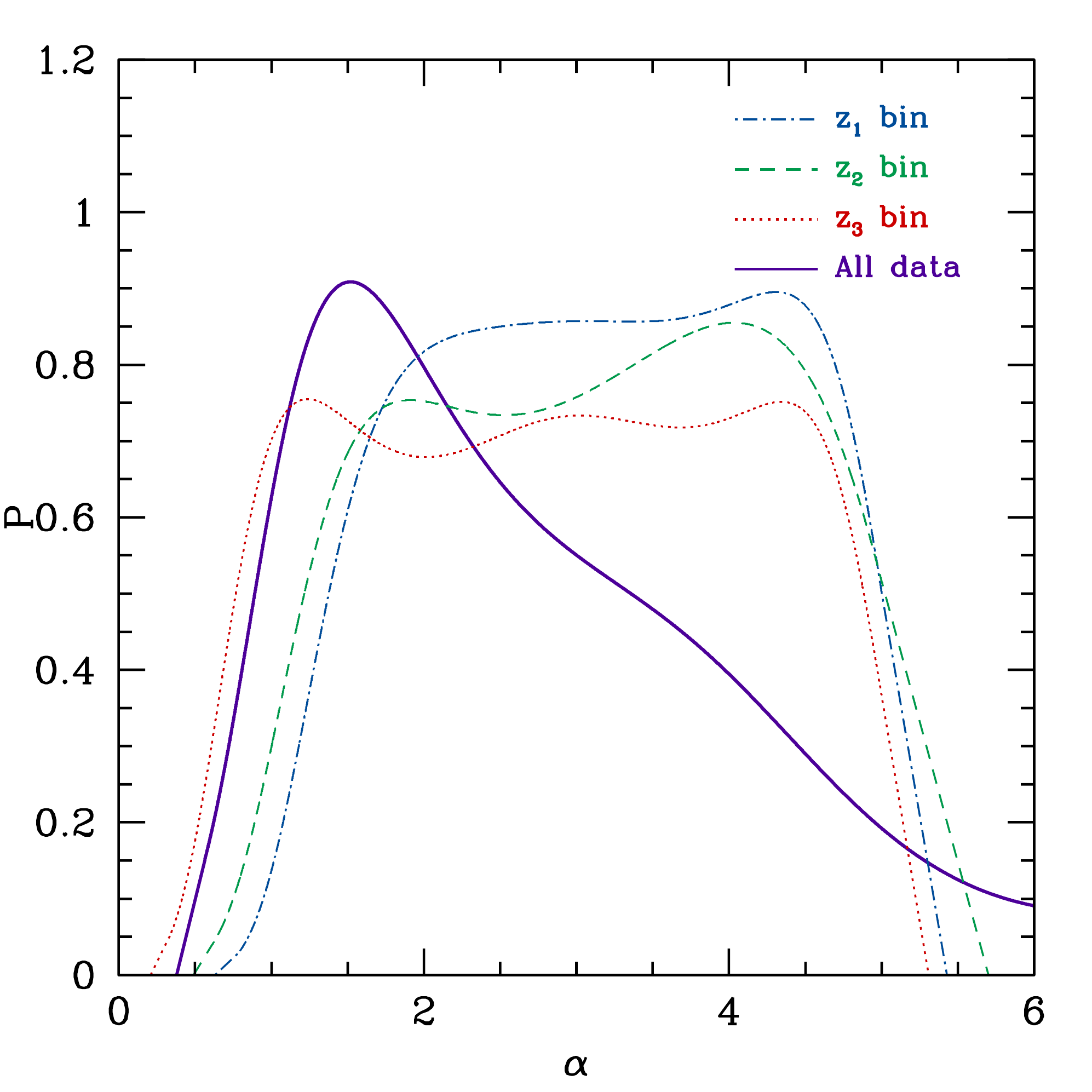}}
\resizebox{!}{!}{\includegraphics[scale=0.41]{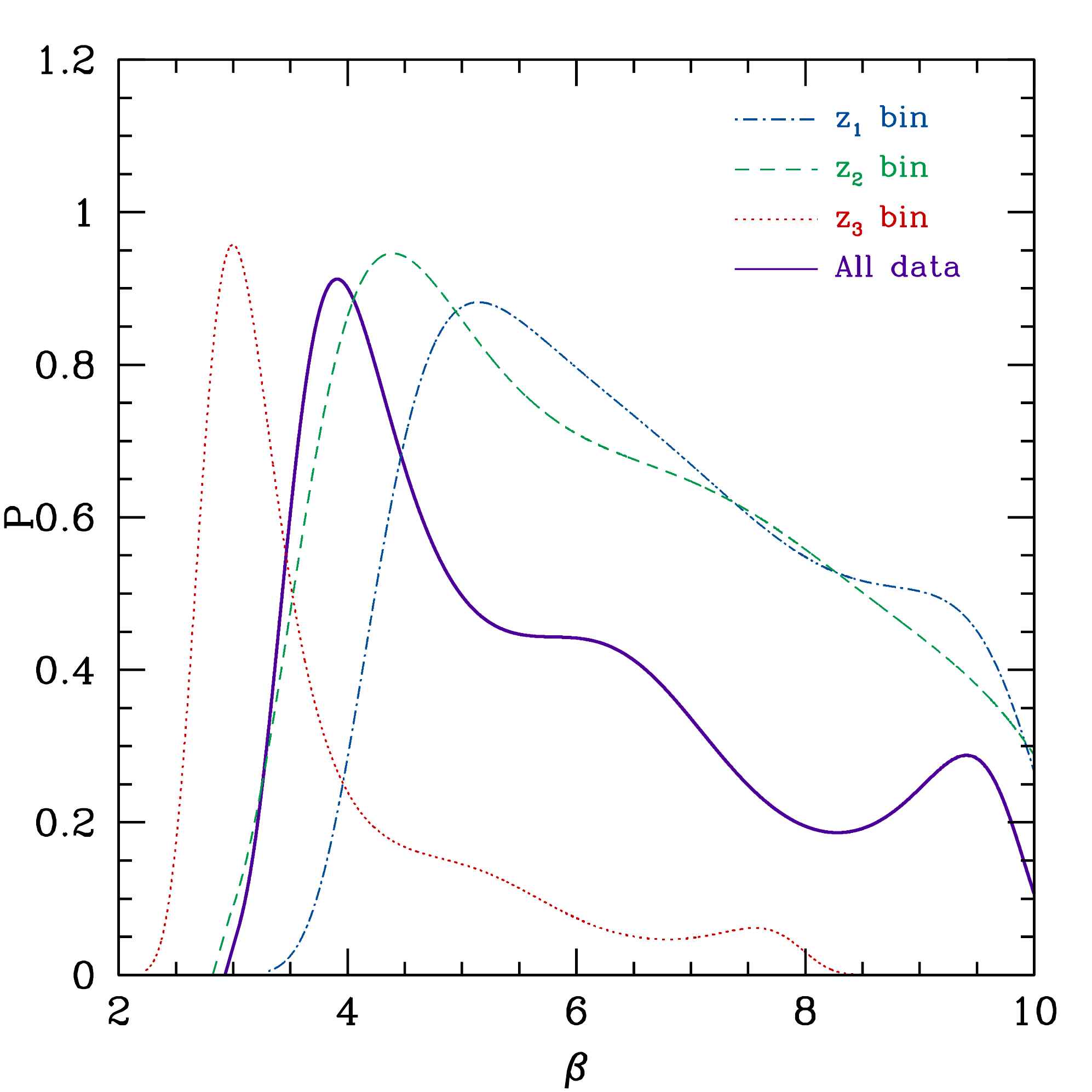}}
}
\caption{\label{fig:1d_pdf} The 1-D PDFs of $P_0$, $c_{500}$, $\alpha$, and $\beta$ in different redshift ranges. The blue dash-dotted, green dashed, and red dotted curves are the results in $z_1$, $z_2$, and $z_3$ bins, respectively. The solid purple curve is for the whole redshift range with all stacked data.}
\end{figure*}

\begin{figure*}
\centerline{
\resizebox{!}{!}{\includegraphics[scale=0.41]{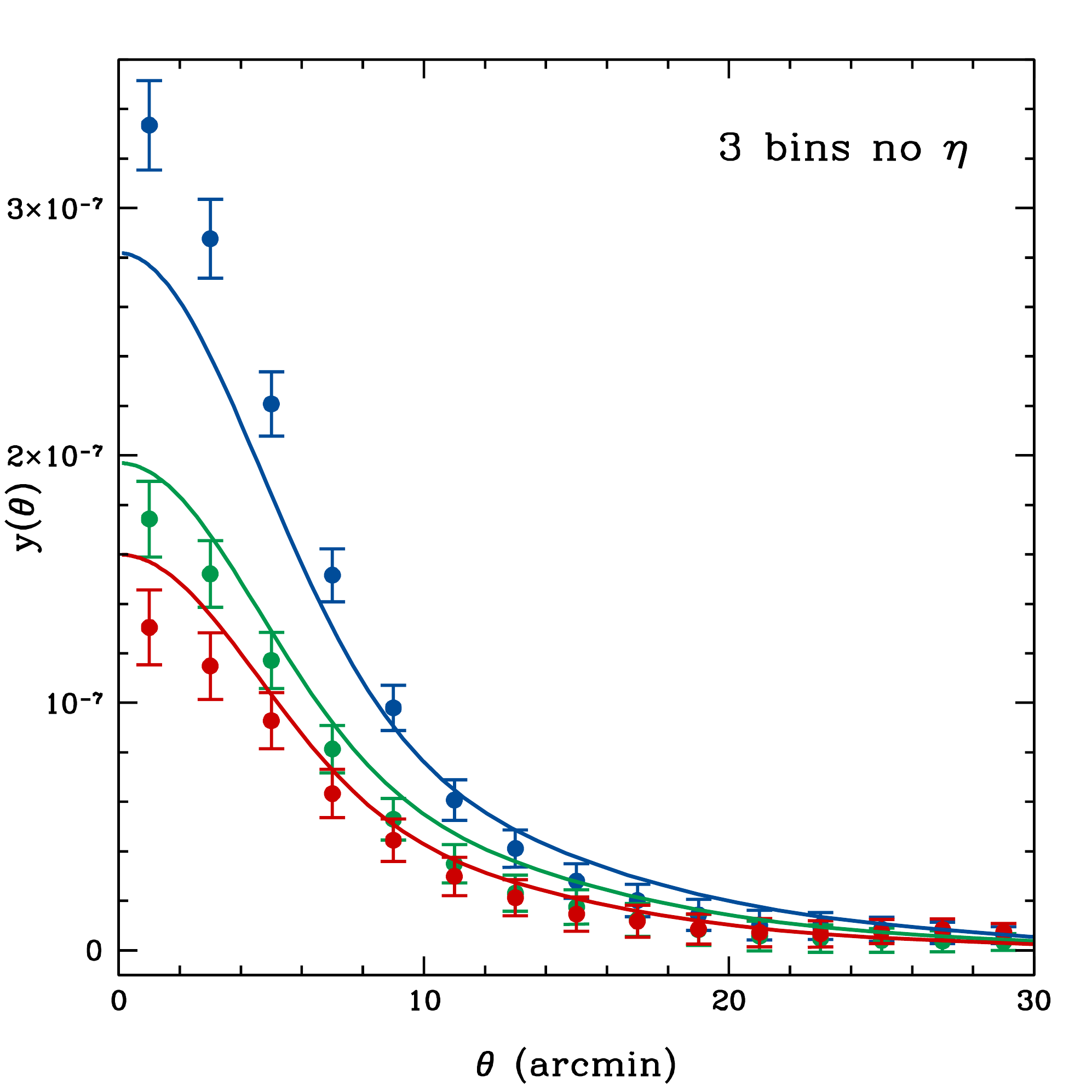}}
\resizebox{!}{!}{\includegraphics[scale=0.41]{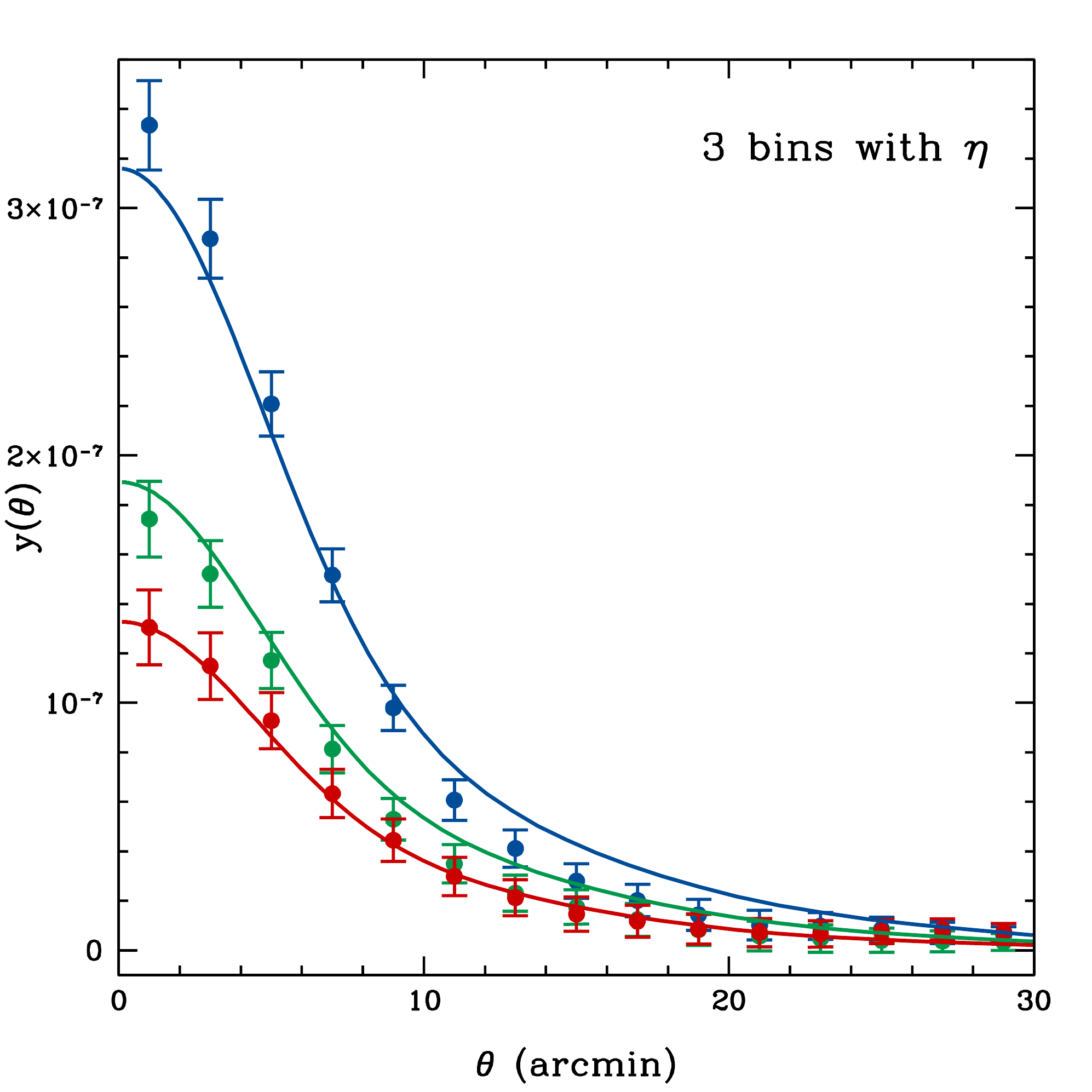}}
}
\caption{\label{fig:3bin_fitting} {\it Left}--The fitting result of $y^{\rm cross}(\theta)$ by simultaneously fitting the three redshift bins without parameter $\eta$. {\it Right}--The same as left panel but $\eta$ included. As can be seen, the data cannot be well fitted within $\theta<5$ arcmin in the $z_1$ and $z_3$ bins in the left panel, and it can be improved by including $\eta$ as shown in the right panel.}
\end{figure*}

\begin{figure*}
\centerline{
\resizebox{!}{!}{\includegraphics[scale=0.8]{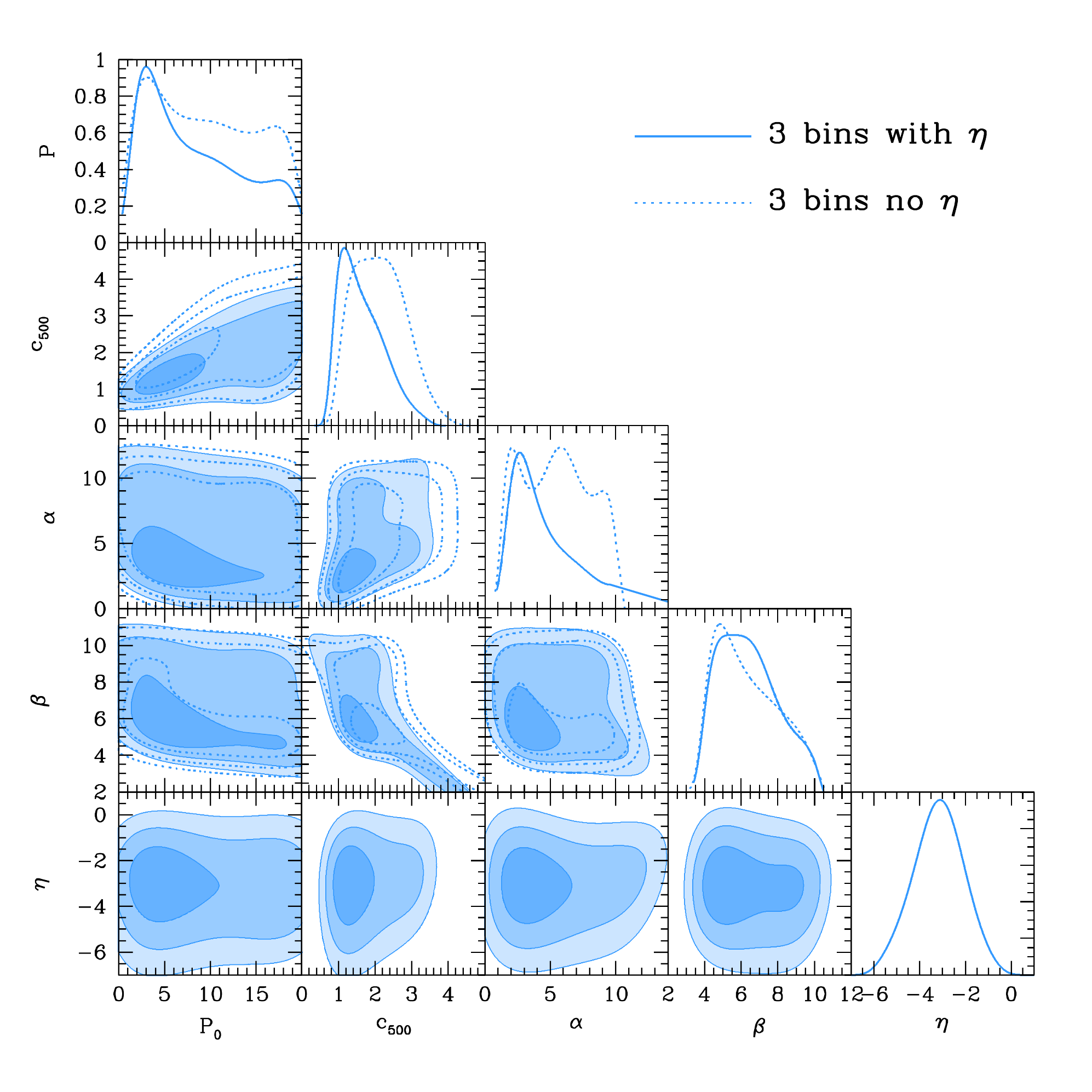}}
}
\caption{\label{fig:contour} The contour maps and 1-D PDF of free parameters when fitting the data of all 3 bins with and without $\eta$. The 68.3\%, 95.5\%, and 99.7\% confidence levels are shown. We can find that the best-fit of $\eta=-3.11^{+1.09}_{-1.13}$, which significantly changes the power-law index $8/3$ of $E(z)$ in Eq.~(\ref{eq:P500}). This indicates that our result prefers weaker redshift-dependence of the electron pressure profile than~\citet{Arnaud10}.}
\end{figure*}

\section{Results of constraints}
\label{sec:result}

In Fig.~\ref{fig:y_profile}, we show the fitting results for different redshift ranges. The solid, dotted, and dashed curves are the best-fits of total, one-halo, and two-halo correlation functions, respectively. Note that we use the covariance matrix between different angular scales instead of the error bars shown in Fig.~\ref{fig:y_profile} in the fitting process. These error bars are simply derived from the diagonal elements of the corresponding covariance matrix. As shown in Table~\ref{tab:fit}, We find that the minimum reduced chi-square, which is defined by $\chi^2_{\rm red}=\chi_{\rm min}^2/(N-M)$ where $N$ and $M$ are the number of data and free parameters in the model, for the three main redshift bins and the whole range with all LRG data are 1.65, 0.45, 0.52, and 1.88, respectively. This indicates that the data can be well fitted in each case, expecially for the $z_2$ and $z_3$ bins.

The corresponding best-fits and 1 $\sigma$ errors of $P_0$, $c_{500}$, $\alpha$, and $\beta$ are shown in Table~\ref{tab:fit}, and the 1-D PDFs are given in Fig.~\ref{fig:1d_pdf}. We find that the $P_0$ is not well constrained by the data, and its 1-D PDFs extend in large parameter space with a wide peak between 2 and 11 in different redshift ranges. The constraint result of $c_{500}$ in each case is in a good agreement that the probability peaks are around 1.5, although the width of the PDF of the $z_3$ bin is wider than others. We don't obtain stringent constraints on $\alpha$ in the three $z$-bins separately. As can be seen, the PDFs have flat tops extending from 1.5 to 4.5 for the $z_1$, $z_2$, and $z_3$ bins. The constraint is significantly improved for the all-data case with a peak at 1.5. For $\beta$, we find that the results of the $z_2$ and $z_3$ bins are well consistent, while the best-fit value is apparently smaller in the $z_3$ bin.

We also find that, generally, the fitting results of the four cases are consistent with that given by \cite{Arnaud10} and \cite{Planck13}, especially for the all-data case (e.g. the constraints on $c_{500}$, $\alpha$, and $\beta$). This implies that our method is feasible and effective for the studies of the cluster electron pressure profile. Beside, as indicated in Table~\ref{tab:fit}, there is noticeable evolution of the parameters, i.e. $c_{500}$, $\alpha$, and $\beta$, in the three redshift bins. We can see that $\alpha$, and $\beta$ become smaller and smaller as the redshift increases, while $c_{500}$ tends to be larger at high redshift.

In order to suppress the fitting uncertainties of the free parameters and check the consistency of the UPP model in different redshift bins, we also try to simultaneously fit the data in the three redshift bins using $\chi^2_{\rm 3bins}=\chi^2_{z_1}+\chi^2_{z_2}+\chi^2_{z_3}$. As shown in the left panel of Fig.~\ref{fig:3bin_fitting}, we find that the data can be fitted with $\chi^2_{\rm red}=1.61$ (see Table~\ref{tab:fit}), adopting the usual model of electron UPP with four free parameters, i.e. $P_0$, $c_{500}$, $\alpha$, and $\beta$\footnote{We also check the fitting result by seting $\gamma$ as a free parameter, and the result is almost the same (but with wider PDFs).}. However, we can see that, except for the $z_2$ bin, the theoretical curves cannot fit the data very well within angular scales less than 5 arcmin in the $z_1$ and $z_3$ bins.

Since we can find that, by comparing to the data, the predicated curve is lower in the $z_1$ bin, and higher in the $z_3$ bin, we can adjust the redshift-dependence of the model by adding a free parameter $\eta$ on the power-law index of $E(z)$ in Eq.~(\ref{eq:P500}) (i. e. $E(z)^{8/3}\to E(z)^{8/3+\eta}$). In the right panel of Fig.~\ref{fig:3bin_fitting}, we show the fitting result when including $\eta$. We can find that the fitting results are significantly improved with $\chi^2_{\rm red}=1.39$, that is $\sim 5$ smaller for the $\chi^2_{\rm min}$ than the case without $\eta$ (see Table~\ref{tab:fit}). We find that the best-fit of $\eta$ is $-3.11^{+1.09}_{-1.13}$, which significantly changes the previous power-law index (i.e. $8/3$) of $E(z)$. This indicates that our result prefers weaker redshift-dependence for the cluster gas pressure model.

The 1-D PDFs and 2-D contour maps of the free parameters for the 3 bins with and without $\eta$ are shown in Fig.~\ref{fig:contour}. The 1$\sigma$ (68.3\%), 2$\sigma$ (95.5\%), and 3$\sigma$ (99.7\%) confidence levels are shown here. We can find that the fitting results of the 3 bins with and without $\eta$ are basically consistent with each other. Besides, by comparing to Figure~5 in \cite{Planck13}, we find that our constraint results (contours and 1-D PDFs) are in a good agreement with their results, and our method can even offer more stringent constraints on $c_{500}$ and $\beta$. The constraint can be further improved in the future with more LRG sample included.

\section{Summary and discussion}
\label{sec:conclude}

In this work, we make use of the cross-correlation between the tSZ signal measured by {\it Planck} satellite and LRGs from SDSS DR7 to study the universal pressure profile of galaxy clusters. We first stack the {\it Planck} $y$-map against the LRGs, and derive the mean $y$ profile. The LRG sample is given by SDSS DR7 with $0.16<z<0.47$ and $3\times10^{12}\lesssim M_{500}\lesssim3\times10^{14}$ $\rm \msun$. In order to study the redshift evolution of the properties of intra-cluster gas, we divide the LRG sample into sub-samples in three redshift bins. We find that the peaks of the mass distributions in the three redshift bins move towards higher halo mass as the redshift increases. In order to remove the effect of non-matching mass distribution in our analysis, we select sub-sample from each redshift bin with the same mass distribution.

Then we derive theoretical stacked tSZ $y$ profile with the help of the halo model. The {\it Planck} beam function is also considered in the estimates. To obtain more realistic and accurate predictions, we take into account of the information from measurements, such as the redshift, $M_{500}$, and $R_{500}$, and simplify the calculation by dividing the LRG sample into sub-redshift and mass bins. We adopt the MCMC technique to illustrate the probability distribution of the parameters in the gas pressure profile, and set wide parameter ranges as prior distributions. 

We separately fit the $y$ profile data obtained from the stacked $y$-map in the $z_1$, $z_2$, and $z_3$ redshift bins, and the whole redshift range with all stacking data. We find that the fitting results of the four parameters, i.e. $P_0$, $c_{500}$, $\alpha$, and $\beta$, are mainly consistent with one another. They are also in a good agreement with the results from \cite{Arnaud10} and \cite{Planck13}, especially for the all-data case.  We find that there is evolution for $c_{500}$, $\alpha$, and $\beta$ from the low to high redshift, that the best-fits of $\alpha$, and $\beta$ become smaller, while $c_{500}$ becomes greater as the redshift increases.

In order to further investigate the redshift evolution and check the consistence of the UPP model at different redshift bins, we fit the data in the three redshift bins simultaneously by summing up the $\chi^2$ of the three bins together. Interestingly, we find that the UPP model cannot provide good fits on the data at $\theta<5$ arcmin in the $z_1$ and $z_3$ bins. After checking the results, we propose to add a parameter $\eta$ on the power-law index of $E(z)$ to change the redshift-dependence of the model. The best-fitting value of $\eta$ is $-3.11^{+1.09}_{-1.13}$, which suggests that, the power index of redshift evolution of the UPP profile should be equal to $\eta +8/3=-0.44^{+1.09}_{-1.13}$. This result implies that the UPP profile may be less redshift-dependent than its original form. Physically, this result indicates that the cluster pressure profile is less evolved than it was thought to be, and Compton $y$ profile could be nearly redshift-independent. By including this factor, we find the $\chi^2_{\rm red}$ decreases from 1.61 to 1.39, which is $\sim5$ smaller in $\chi^2_{\rm min}$ than that given by the usual UPP model. By comparing the 1-D and 2-D PDFs with \cite{Planck13}, we find our results from the 3 bins with and without $\eta$ cases can match theirs very well, and can even offer more stringent constraints on $c_{500}$ and $\beta$. This indicates that our method can provide reliable results which prefers less redshift-dependence of the UPP mode. We will include more samples and further confirm our result in the future work, e.g. analyzing the SDSS DR12 data, and provide more accurate constraints on the cluster pressure profile.

Besides, this study also can be an important step towards fully quantify the distribution of missing baryons. This is because, a significant amount of baryons are associated with filaments, voids and sheets which have much weaker signals of SZ effect than halos~\citep{Haider16}. As one can see in~\citet{Tanimura19a}, the stacked SZ signal of the filaments is usually entangled with the halo contribution. Therefore, improvement on the halo model's pressure profile will lead to a more precise subtraction of the halo contribution, and will result in better measurement of the signals from filaments and sheets. Although this is out of the scope of this paper, our study can eventually contribute to the more precise determination of the baryons within filaments and sheets.

\section*{Acknowledgements}
YG acknowledges the support of NSFC-11822305, NSFC-11773031, NSFC-11633004, the Chinese Academy of Sciences (CAS) Strategic Priority Research Program XDA15020200, the NSFC-ISF joint research program No. 11761141012, and CAS Interdisciplinary Innovation Team.  Y.Z.M. is supported by the National Research Foundation of South Africa with Grant no.105925, no. 110984, and NSFC-11828301. This research has been also supported by funds from the European Research Council (ERC) under the Horizon 2020 research and innovation programme grant agreement of the European Union: ERC-2015-AdG 695561 (ByoPiC, https://byopic.eu).

\appendix
\section{Stacked \lowercase{t}SZ profile in physical scales}

\begin{figure*}
\centerline{
\resizebox{!}{!}{\includegraphics[scale=0.4]{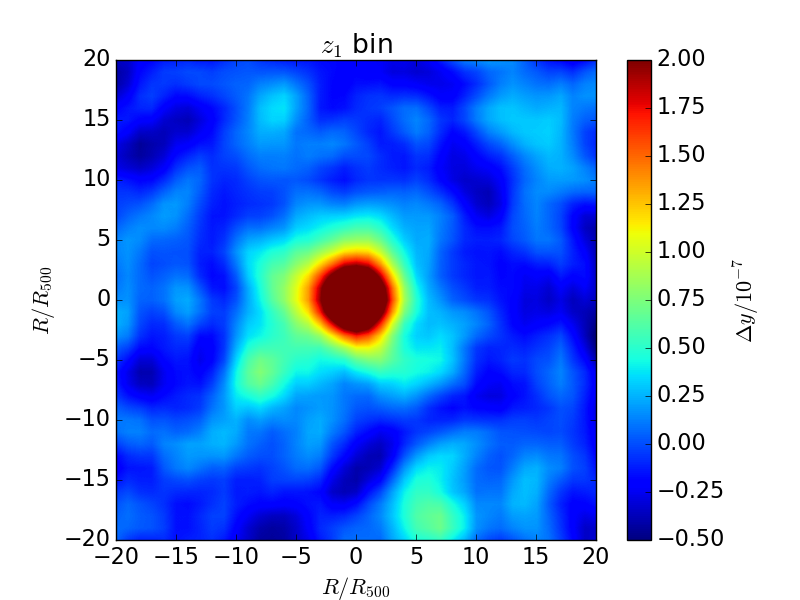}}
\resizebox{!}{!}{\includegraphics[scale=0.4]{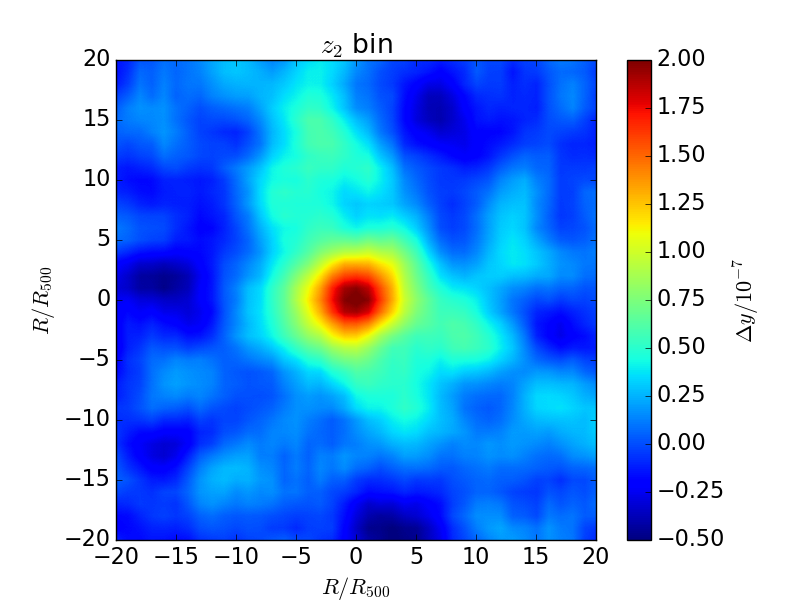}}
}
\centerline{
\resizebox{!}{!}{\includegraphics[scale=0.4]{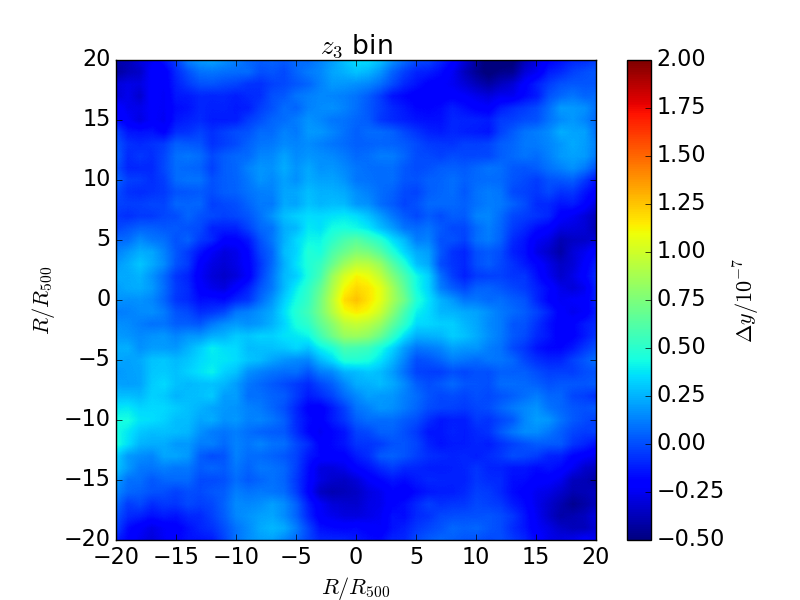}}
\resizebox{!}{!}{\includegraphics[scale=0.4]{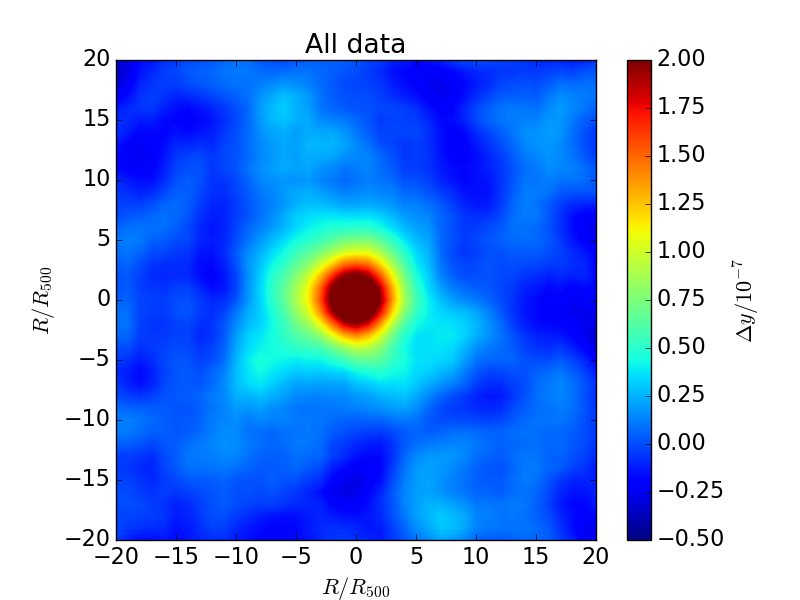}}
}
\caption{\label{fig:ymap_R500} The stacked $y$ intensity maps for $z_1$ bin (upper left), $z_2$ bin (upper right), $z_3$ bin (bottom left), and the whole redshift range with all LRG data (bottom right). The number of selected LRGs stacked in each map is 11,926 with the same mass distribution. The mean tSZ intensity in the annular region between $r/R_{500}=$15 and 20 has been subtracted as the local background.}
\end{figure*}

\begin{figure*}
\centerline{
\resizebox{!}{!}{\includegraphics[scale=0.4]{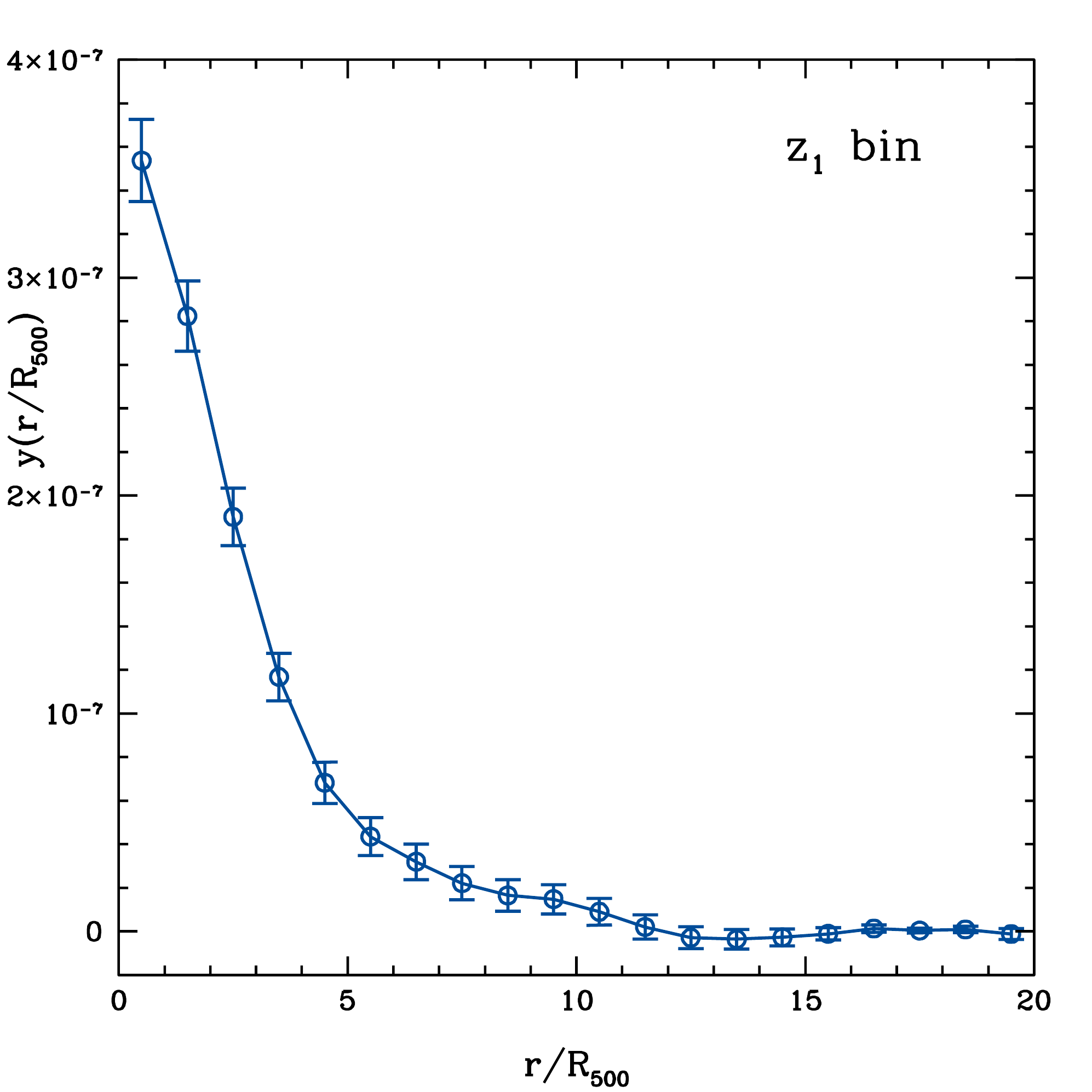}}
\resizebox{!}{!}{\includegraphics[scale=0.4]{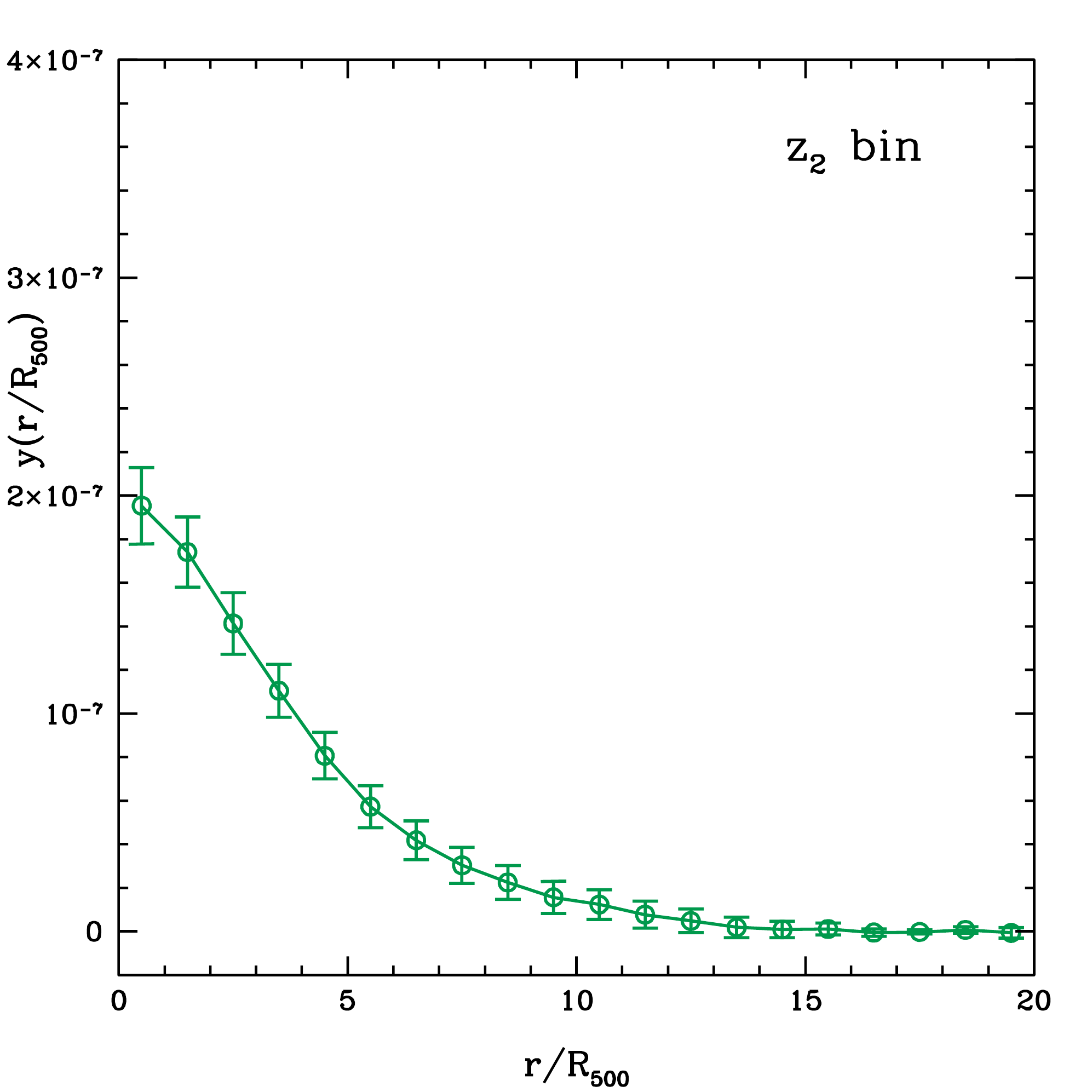}}
}
\centerline{
\resizebox{!}{!}{\includegraphics[scale=0.4]{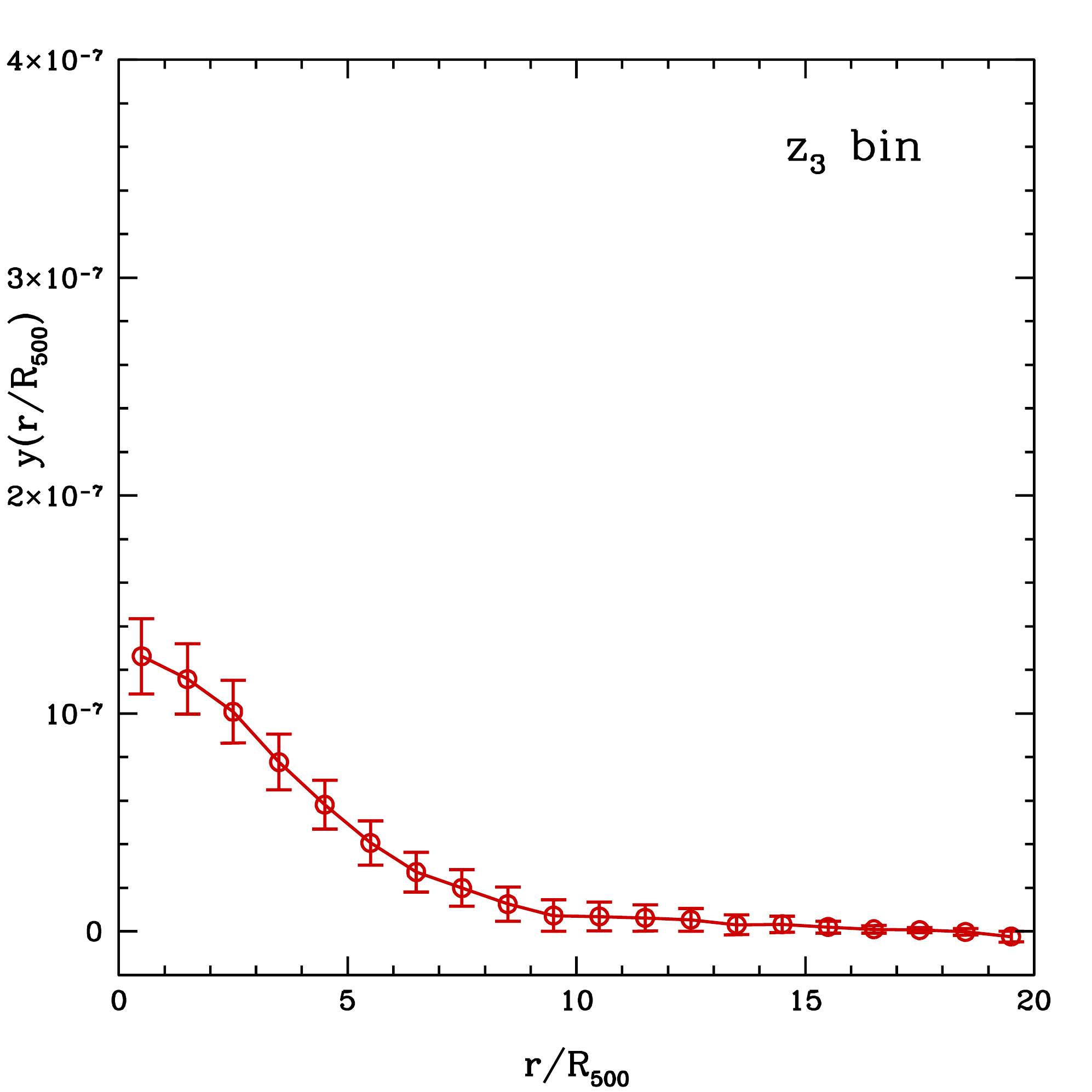}}
\resizebox{!}{!}{\includegraphics[scale=0.4]{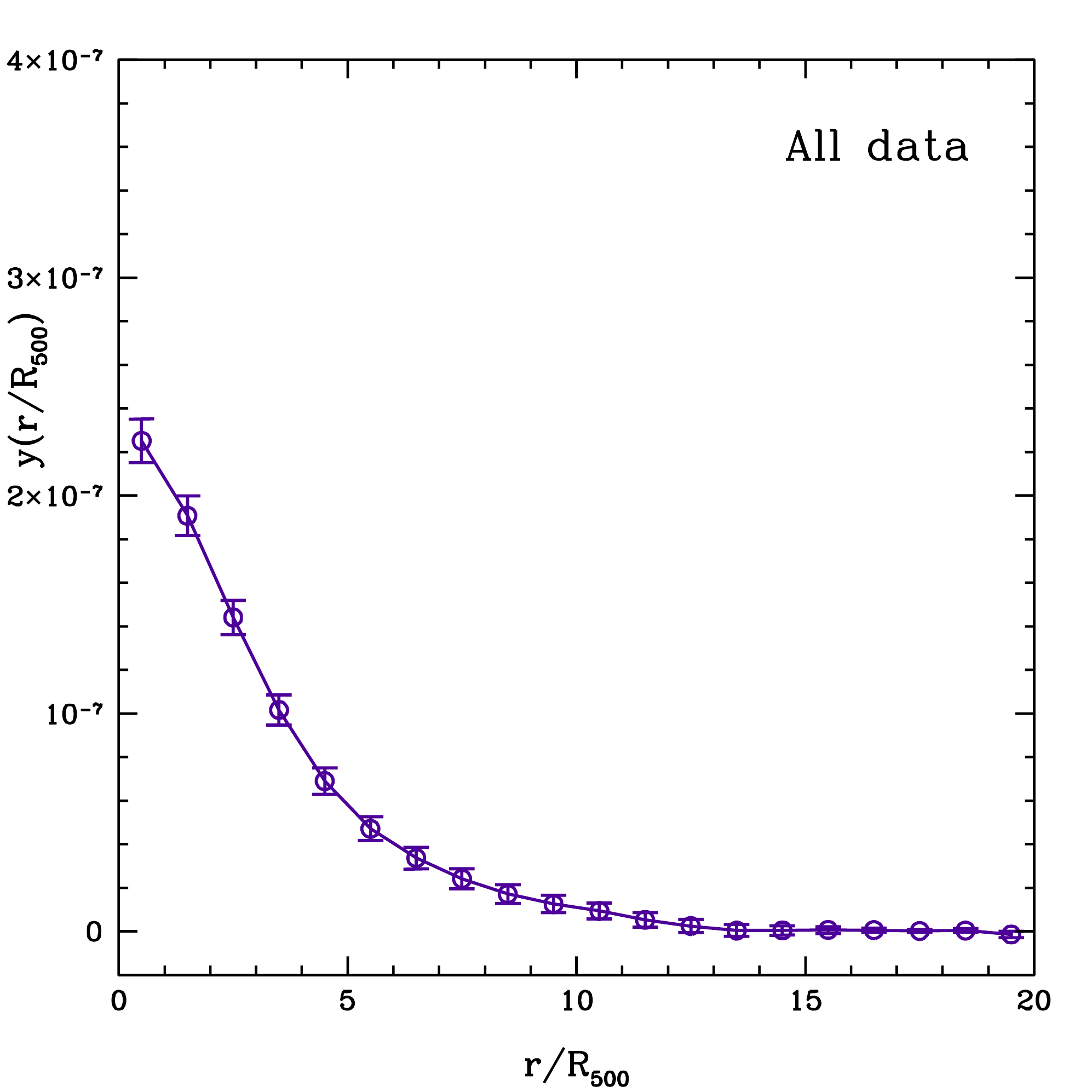}}
}
\caption{\label{fig:y_profile_R500} The cross correlations of tSZ signal and galaxy cluster distribution as a function of $r/R_{500}$ in different redshift ranges. The error bars are simply derived from the diagonal elements of the covariance matrix in each case.  All of data points are rescaled based on the average values of corresponding intensities between $r/R_{500}=$15 and 20.}
\end{figure*}

\begin{figure}
\centerline{
\resizebox{!}{!}{\includegraphics[scale=0.33]{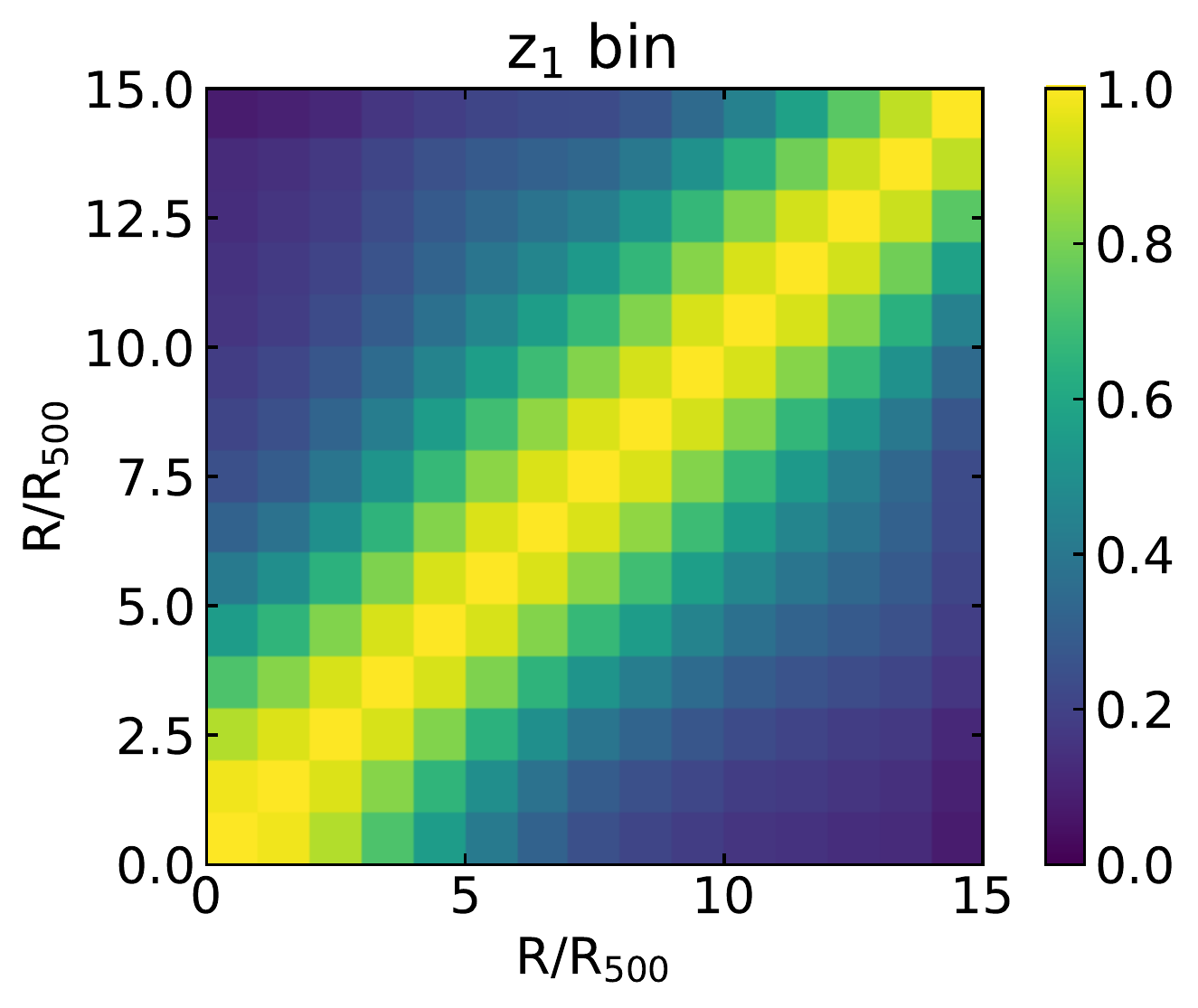}}
\resizebox{!}{!}{\includegraphics[scale=0.33]{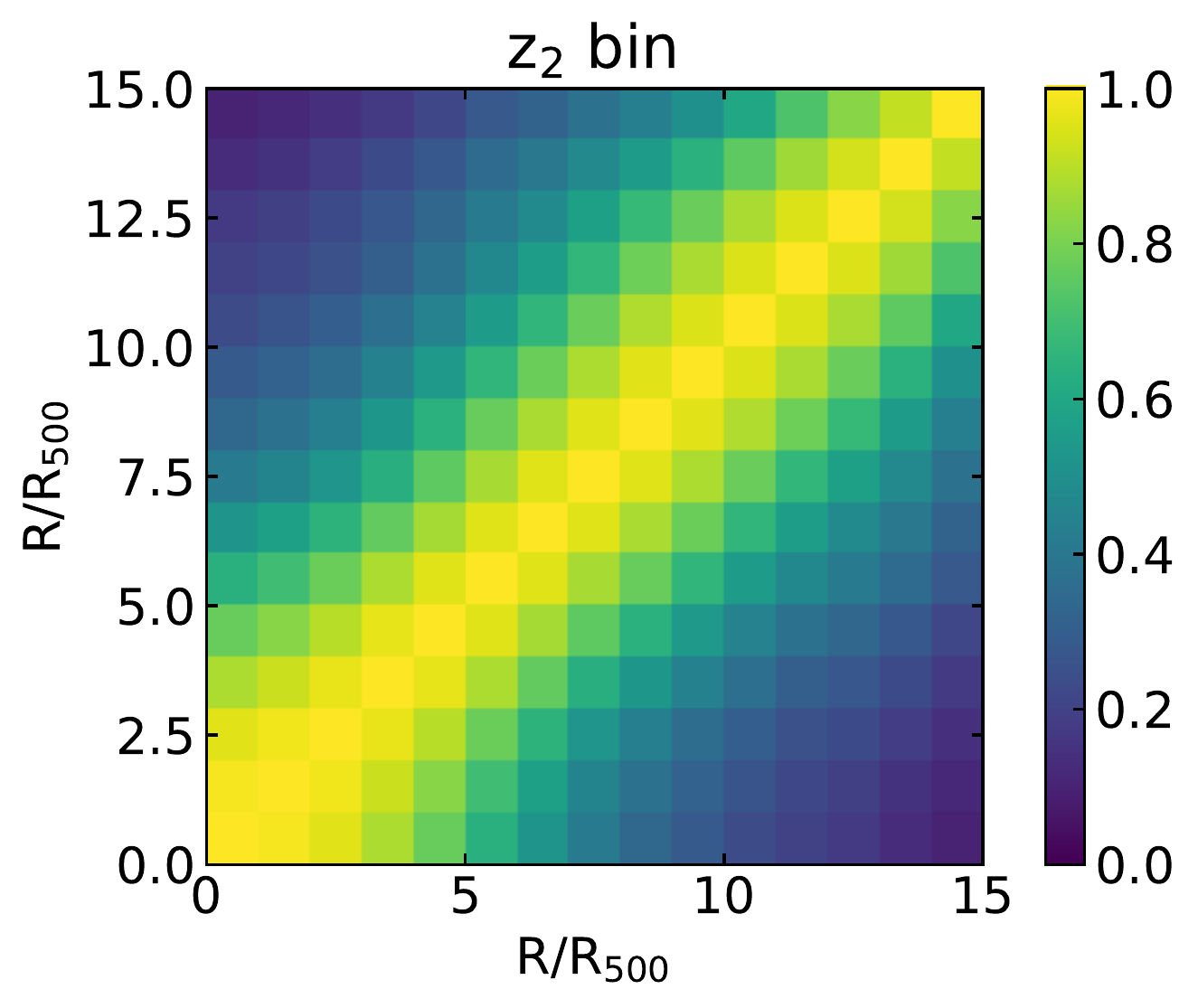}}
}
\centerline{
\resizebox{!}{!}{\includegraphics[scale=0.33]{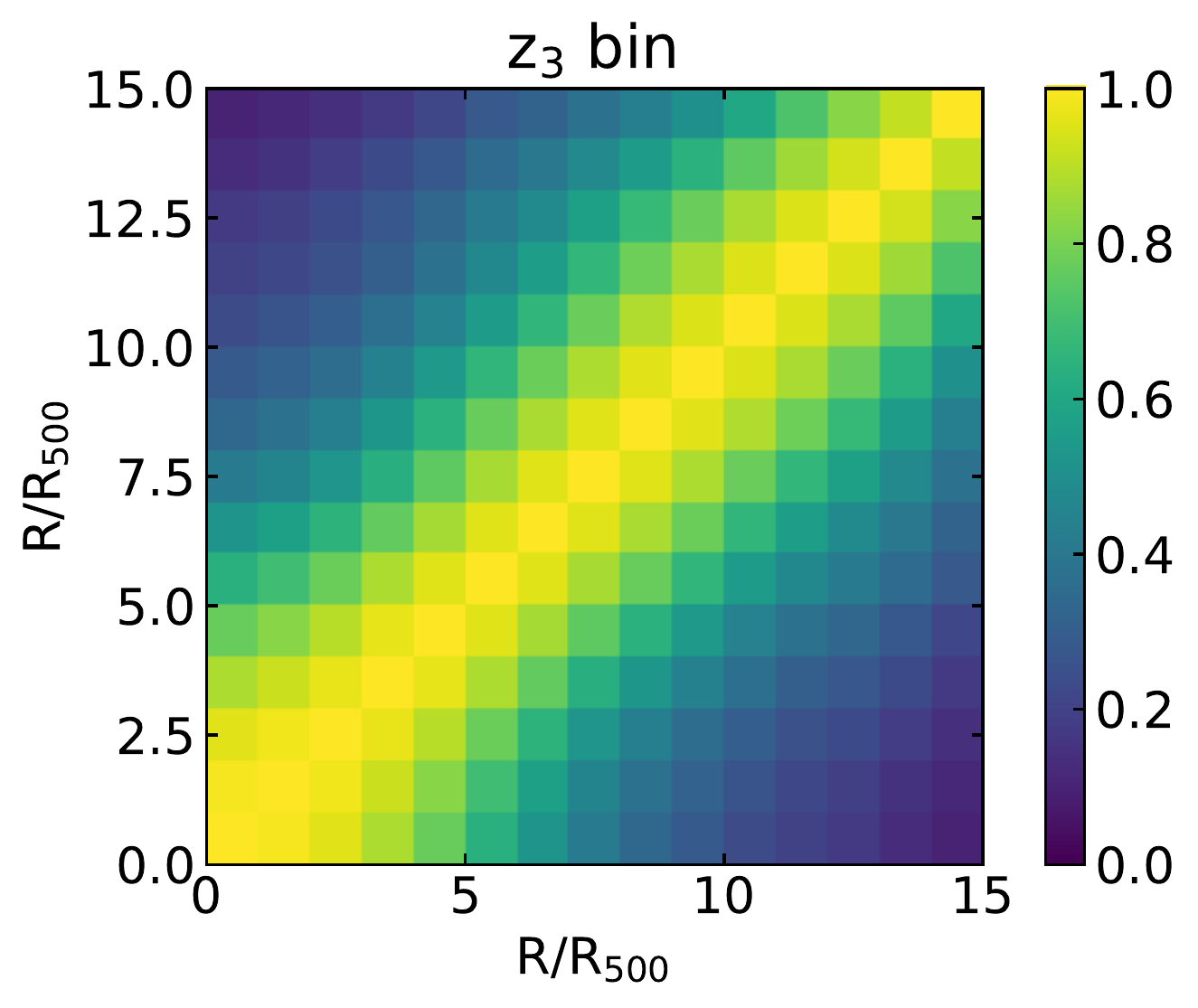}}
\resizebox{!}{!}{\includegraphics[scale=0.33]{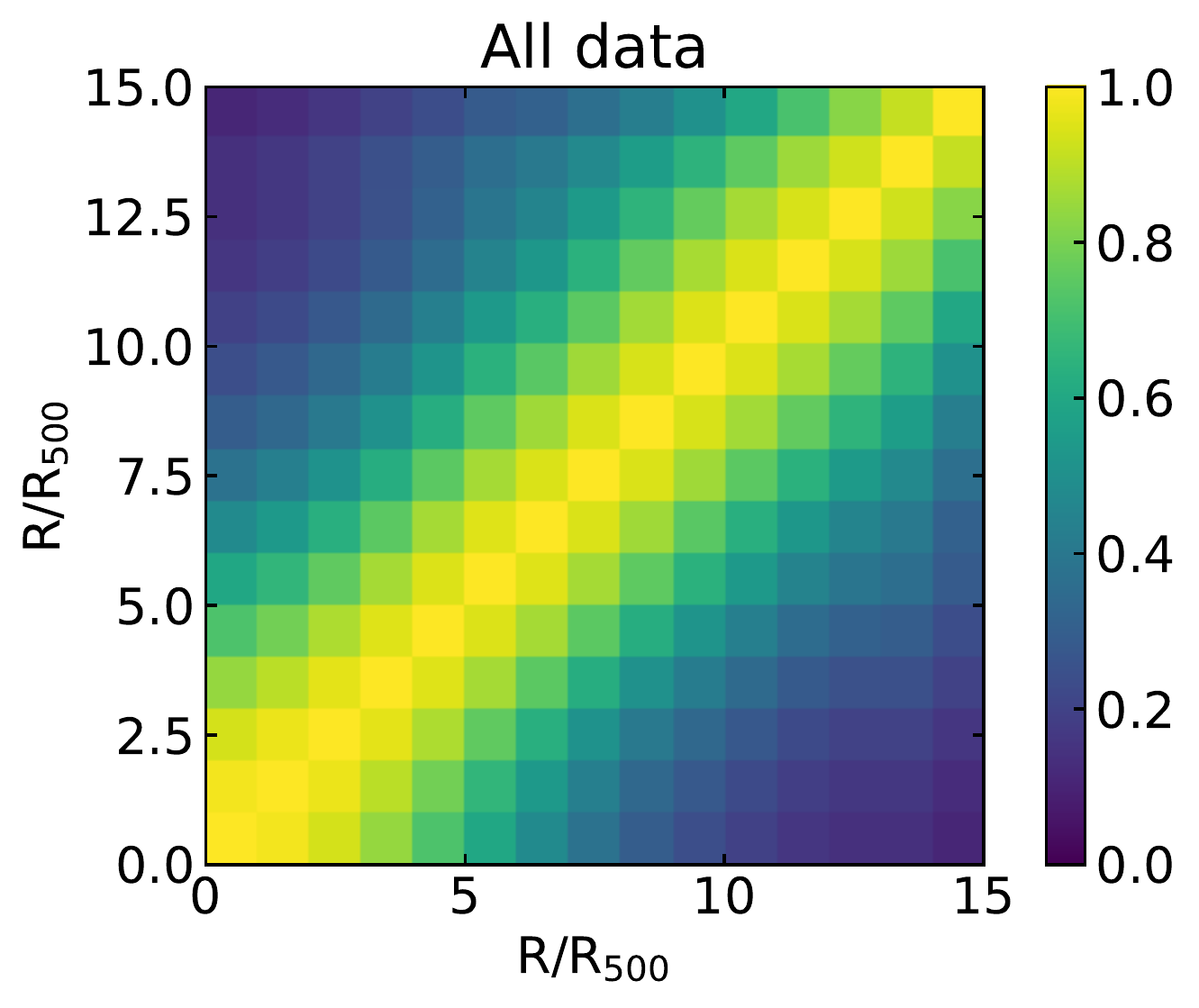}}
}
\caption{\label{fig:coeff_R500} The correlation coefficient matrices between physical scales for the $y$ profiles $y(r/R_{500})$ in different redshift ranges. The scale $r/R_{500}$ has been divided into 15 bins.}
\end{figure}

In order to stack the tSZ signal at the same physical scales from each LRG, we stack the \Planck $y$-map at the positions of LRGs in physical coordinates, instead of angular coordinate. We reset a 2-dimensional physical coordinate system of $-20 < r/R_{500} < 20$ and $-20 < r/R_{500} < 20$, divided in 80 $\times$ 80 bins. The local background region is also redefined in physical scale, which is an annular region between $|r/R_{500}|=$15 and 20  for each LRG. Then, we follow the same procedure as described in Sect.~\ref{subsec:stacked-map} for stacking. After removing the LRGs with $\ge20\%$ masked region within a $r/R_{500}=20$ circle, we obtain 18,117, 29,074, and 26,401 LRGs, respectively, for the $z_1$, $z_2$, $z_3$ bins, and total 73,592 in the whole redshift range. We finally select 11,926 LRGs in each redshift bin with the same mass distribution. This mass distribution is quite similar with that in the $y(\theta)$ case. 

The $y$-maps, cross correlations of tSZ signal and galaxy cluster distribution, and correlation coefficient matrices between physical scales in the $y(r/R_{500})$ case are shown in Figure~\ref{fig:ymap_R500}, \ref{fig:y_profile_R500}, \ref{fig:coeff_R500}, respectively. By comparing to the $y(\theta)$ case, as shown in Figure~\ref{fig:y_profile} and \ref{fig:y_profile_R500}, we find that there is a bump feature around $r/R_{500}\sim10$ in the $y(r/R_{500})$ profiles, especially for the $z_1$ bin, which is mainly due to the two-halo term. This feature is smoothed out in the $y(\theta)$, since it is obtained by stacking different cluster physical scales at the same angular scale $\theta$.









\bsp	
\label{lastpage}
\end{document}